\documentclass[aps,pra,superscriptaddress,reprint,onecolumn,notitlepage,nofootinbib]{revtex4-1}

\usepackage{usepkg}
\usepackage{math}

\newcommand{\bE}{\mathbb{E}}

\definecolor{darkred}{RGB}{200, 0, 0}
\definecolor{darkblue}{RGB}{0, 0, 180}
\definecolor{darkgreen}{RGB}{50, 150, 0}
\definecolor{col2}{HTML}{64A857}
\definecolor{col3}{HTML}{D1603D}
\definecolor{orange}{rgb}{0.9,0,0}

\begin{document}

\title{On Bit Commitment and Oblivious Transfer in Measurement-Device Independent settings}

\author{J\'{e}r\'{e}my Ribeiro}
\affiliation{QuTech, Delft University of Technology, Lorentzweg 1, 2628 CJ Delft, The Netherlands}
\affiliation{Kavli Institute of Nanoscience, Delft University of Technology, Lorentzweg 1, 2628 CJ Delft, The Netherlands}

\author{Stephanie Wehner}
\affiliation{QuTech, Delft University of Technology, Lorentzweg 1, 2628 CJ Delft, The Netherlands}
\affiliation{Kavli Institute of Nanoscience, Delft University of Technology, Lorentzweg 1, 2628 CJ Delft, The Netherlands}

\date{\today}

\begin{abstract}
    Among the most studied tasks in Quantum Cryptography one can find Bit Commitment (BC)
   and Oblivious Transfer (OT), two central cryptographic primitives.
   In this paper we propose for the first time protocols for these tasks in the measurement-device
   independent (MDI) settings and analyze their security.
   We analyze two different cases: first we assume the parties
   have access to perfect single photon sources (but still in the presence of noise and losses), and second
   we assume that they only have imperfect single photon sources.
   In the first case we propose a protocol for both BC and OT and prove their security in the
   Noisy Quantum Storage model.
   Interestingly, in the case where honest parties do \emph{not} have access to perfect single photon sources,
   we find that BC is still possible, but that it is ``more difficult'' to get a secure protocol for OT:
   We show that there is a whole class of protocols that cannot be secure. All our
   security analyses are done in the finite round regime.

\end{abstract}

\maketitle

\section{Introduction}

Oblivious Transfer (OT) -- \cite{Rabin81} -- and Bit Commitment
(BC) --\cite{GMW91,Naor89,CDG87} -- are two central, and well studied
cryptographic primitives. In fact it has been shown that OT is universal \cite{Kilian} in the sense that
all two-partite secure function evaluation tasks can be
reduced to OT. This means that if one is given a secure implementation of OT, one can construct a protocol
using OT (and classical communications) that implements any secure function evaluation protocol.

OT and BC are related to each other. In particular since OT is universal,
it is possible to implement BC given an OT routine. The converse is not true if
we limit the parties to classical communication \cite{MN05}. However if the parties
have access to quantum communication it can be shown that OT can be reduced to
BC \cite{Crepeau92,FS09}, and therefore OT and BC are equivalent in the quantum settings.

Unfortunately it is now well known that neither OT nor BC can be implemented when
no restriction (other than following the laws of (quantum) physics) is made on
the power of the adversary \cite{Mayers1997, LoChau97, Lo_whyquantum}. This
motivated the search for realistic assumptions that could be made on the adversary's
power.

Inspired by the classical Bounded Storage Model \cite{Maurer92b},
Ref.~\cite{Damgard05} proposed protocols assuming that the adversary can only store a limited amount
of qubits, \emph{i.e.}~that the quantum memory of the adversary is bounded. The assumption is
called the Bounded Quantum Storage Model. Note that the Bounded Quantum Storage Model has the
advantage over computational assumptions that it allows for everlasting security, meaning that
the assumption only needs to be satisfied \emph{during} the execution of the protocol. No
additional resources or power gained after the execution of the protocol can allow the
adversary to break security. This contrasts with computational assumption for which
giving more computational power to the adversary after the execution of the protocol is a threat to the security of the protocol.
The Bounded Quantum Storage Model -- as well as the more general Noisy
Quantum Storage Model -- allows to prove the security of protocols for
OT and BC \cite{Damgard05,Damgard2007,WSEE,Steph_1,NJCKW12,ENGLWW15}.
However, these security proofs rely implicitly on the assumption that the devices
used by the honest parties are {sufficiently well characterized} and will always work as expected. This assumption
might not always be satisfied in practice. In particular, in the context of Quantum Key Distribution (QKD),
attacks performed by tampering with the measurement devices exist ~\cite{Makarov06,Sajeed15}.

In the context of quantum key distribution, in order to present protocols that
are not subject to these types of attack, Refs.~\cite{mayers98,Acin07,PAB09} propose a security proof
that is independent of the inner working of the quantum devices used during the protocol.
In fact the devices are considered as black boxes, and the security
only relies on the ability of the devices to demonstrate certain
``non-local'' statistics for their inputs and outputs. More precisely the authors
show that if the devices are able to violate the CHSH inequality \cite{CHSH69}
then there is a secure protocol for quantum key distribution. This result has
been later generalized to include a more powerful adversary \cite{VV12,MS14,MS14a,ADFORV18,RMW18}.
The model in which the devices are considered as black boxes is called device independence.

Following this idea of device independence Refs.~\cite{KW16,RPKHW18} proved
security of BC and OT in the bounded/noisy quantum storage model in device independent
settings. However it is important to note that the authors assume that, even if the devices may
behave in an arbitrary way, they do so in the same fashion in every use of the devices
independently of the past. In other words they assume that the devices are
memoryless. Other protocols \cite{aharon15,Silman11} are secure against a more powerful adversary
but require different settings where they only achieve an imperfect bit commitment
scheme. In general it is quite hard to prove device independent security
of protocols. In particular there is no known security proof for device independent
OT and BC in the settings presented in Refs.~\cite{KW16,RPKHW18} without the
memoryless assumption. Experimental implementations of device independent protocols
are also a lot more demanding
as discussed in Ref.~\cite{MVRHW18} for quantum key distribution. In fact it is so
demanding that, while many quantum key
distribution and some (quantum) BC protocols have been implemented,
there has not been any device independent implementation of these protocols
so far, not even assuming that the devices are memoryless.

\begin{figure}[!h]
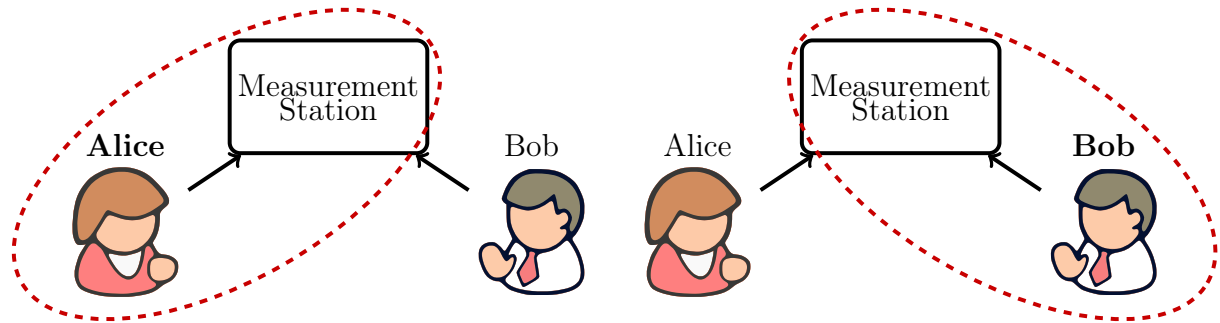

  \begin{subfigure}{0.45\textwidth}
  \begin{tikzpicture}[line width=1.5pt,scale=0.9]
    \node (Alice) at (-3,-1) {\includegraphics[width=40pt]{Alice.pdf}};
    \node[above=25pt] at (Alice){\large {\bf Alice}};
    \node (Bob) at (3,-1) {\includegraphics[width=40pt]{Bob.pdf}};
    \node[above=25pt] at (Bob){\large Bob};
    \node (Station)[draw,rectangle,rounded corners,minimum width=1.5cm,minimum height=1.5cm] at (0,1){$\substack{\text{\large Measurement}\\\text{\large Station}}$};
    \draw[->] (Alice) -- (Station);
    \draw[->] (Bob) -- (Station);
    \draw[color=darkred, dashed] (-1.5,0) circle [x radius=3.5cm, y radius=17mm, rotate=30];
  \end{tikzpicture}
  \caption{Schematic of an MDI protocol with dishonest Alice. Alice has control over the measurement station, therefore
  we will treat Alice and the measurement station as one party.}
  \label{fig:dishonest_Alice}
\end{subfigure}
\begin{subfigure}{0.5\textwidth}
\begin{tikzpicture}[line width=1.5pt,scale=0.9]
  \node (Alice) at (-3,-1) {\includegraphics[width=40pt]{Alice.pdf}};
  \node[above=25pt] at (Alice){\large Alice};
  \node (Bob) at (3,-1) {\includegraphics[width=40pt]{Bob.pdf}};
  \node[above=25pt] at (Bob){\large {\bf Bob}};
  \node (Station)[draw,rectangle,rounded corners,minimum width=1.5cm,minimum height=1.5cm] at (0,1){$\substack{\text{\large Measurement}\\\text{\large Station}}$};
  \draw[->] (Alice) -- (Station);
  \draw[->] (Bob) -- (Station);
  \draw[color=darkred, dashed] (1.5,0) circle [x radius=3.5cm, y radius=17mm, rotate=-30];
\end{tikzpicture}
\caption{Schematic of an MDI protocol with dishonest Bob. \\Bob has control over the measurement station, therefore\\
we will treat Bob and the measurement station as one \\party.}
\label{fig:dishonest_Bob}
\end{subfigure}
\caption{Schematic MDI protocol with dishonest Alice or dishonest Bob.}
\end{figure}

These difficulties together with the fact that many attacks on the non device independent
protocols \cite{Makarov06,Sajeed15} are tampering with the measurement devices and
not with the photon sources (or quantum state sources), has led Refs.\cite{LCQ12} to introduce
a weaker but more practical notion of device independence called measurement-device independence (MDI).
Here only the measurement devices are treated as black boxes, not the sources of photons (states)
that are still trusted. Since then many measurement-device independent protocols have been implemented \cite{LCW13,POS15,TLX14,TYC14,FVX13} {for QKD}.
Typically, in a measurement-device independent protocol, all the measurement devices are
in a measurement station in between the parties (see Fig.~\ref{fig:MDI-setting}). {
Having the measurement station in between the two parties  is
very natural if one considers that it is part of the network infrastructure also used for QKD.}
The parties will send BB84 type states to the measurement station which will
perform a joint Bell measurement on the incoming qubits. As there is no assumption on these devices
we will {here} always assume that the dishonest party can control the station (see Fig.~\ref{fig:dishonest_Alice}\ref{fig:dishonest_Bob}).
This situation is different from MDI QKD, where the dishonest party is
always a third party who only controls the measurement station, but never {the
sources of Alice and Bob}. In particular, in QKD, Alice and Bob can always trust each other, which is not the case
for BC or OT.

\begin{Myfigure}
  \begin{tikzpicture}[line width=1.5pt,scale=1.2]
    \node (Alice) at (-3,-1) {\includegraphics[width=40pt]{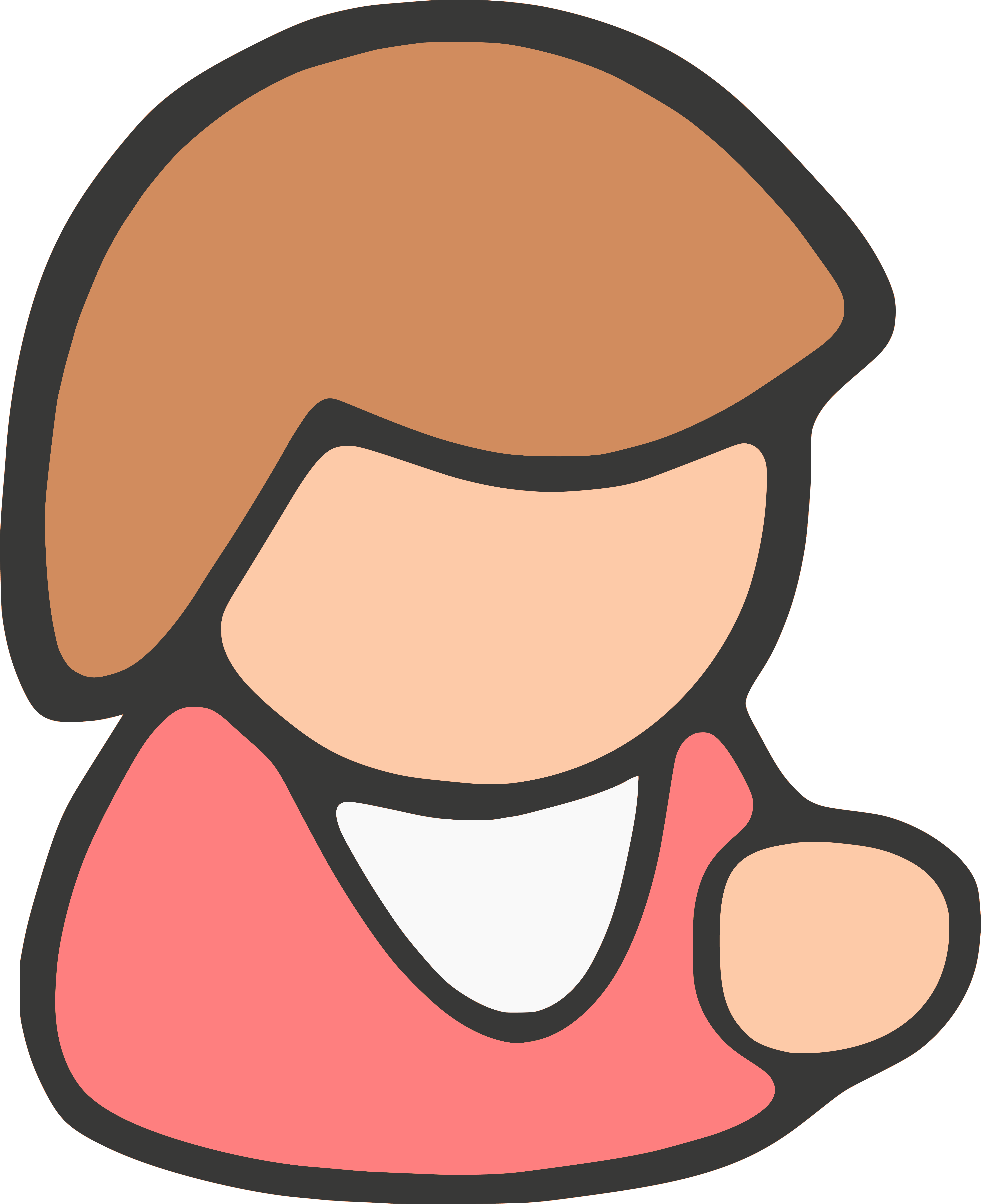}};
    \node[above=25pt] at (Alice){\large Alice};
    \node (Bob) at (3,-1) {\includegraphics[width=40pt]{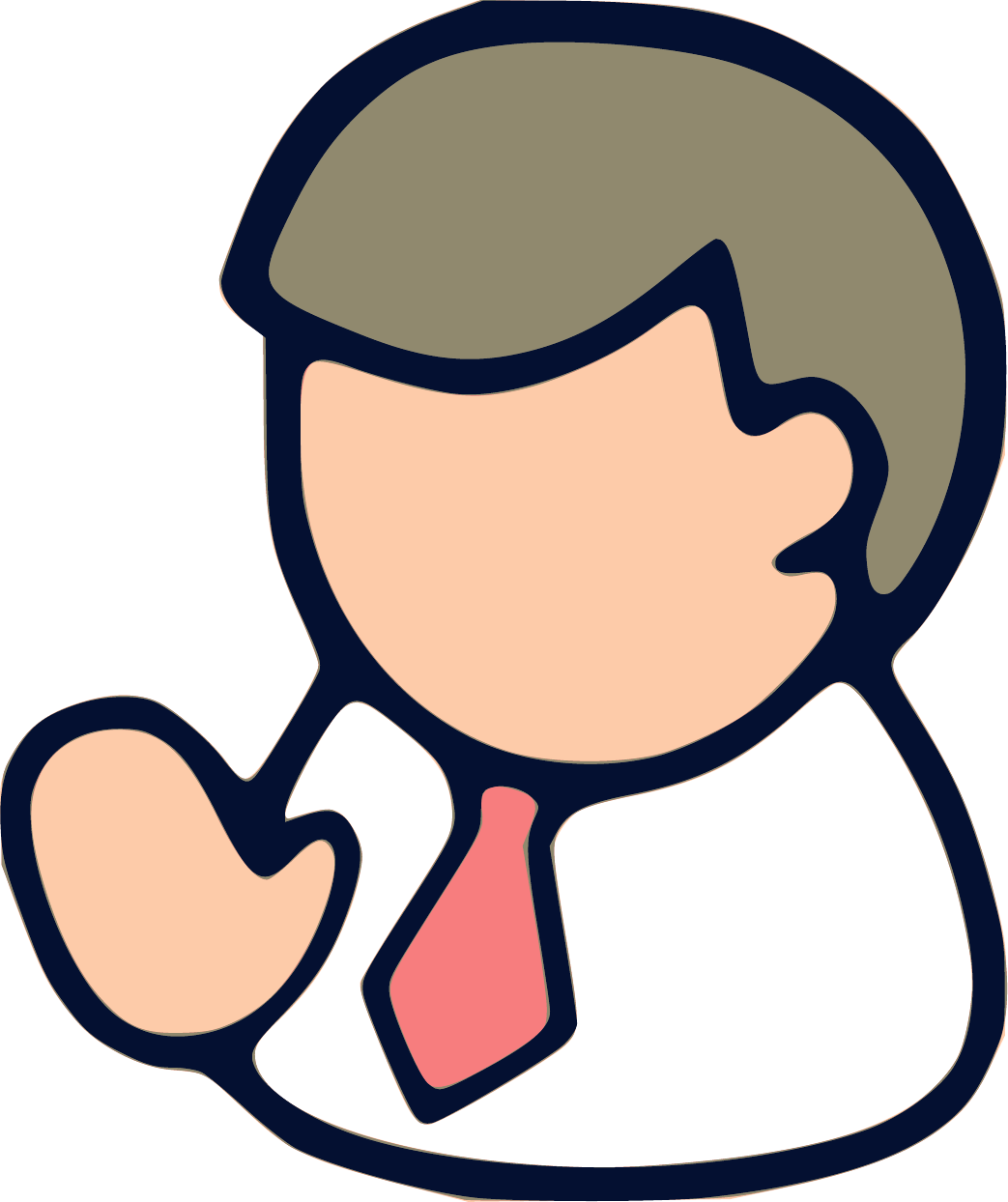}};
    \node[above=25pt] at (Bob){\large Bob};
    \node (Station)[draw,rectangle,rounded corners,minimum width=1.5cm,minimum height=1.5cm] at (0,1){$\substack{\text{\large Measurement}\\\text{\large Station}}$};
    \draw[->] (Alice) -- (Station)node[right=6pt,pos=0.5]{$\ket{\Psi}$};
    \draw[->] (Bob) -- (Station)node[left=6pt,pos=0.5]{$\ket{\Psi'}$};
  \end{tikzpicture}
  \captionof{figure}{Schematic of a MDI protocol.}
  \label{fig:MDI-setting}
\end{Myfigure}

However, almost all the work on measurement-device independence is focused on quantum key distribution \cite{LCQ12,LCW13,POS15,TLX14,TYC14},
and as far as we know there is no proposed protocol for BC or OT in the measurement-device independent
settings. In this work we present protocols for BC and OT and analyze their security. Importantly,
all our security proofs hold in the finite rounds regime and can be implemented with current
state-of-the-art quantum technologies. We first analyze the situation where the
honest parties have perfect single photon sources. Interestingly, in the case where honest parties do not have access to perfect single photon sources,
we find that BC is still possible, but that it is ``more difficult'' to get a secure protocol for OT:
We show that there is a whole class of
protocols that cannot be secure. We present in the next section a
detailed overview of our results.

\subsubsection{Notation}
{In this paper we will denote quantum states by the Greek letters $\rho, \sigma$.
For the purpose of this analysis quantum states can be taken to be
positive linear operators of trace equal to $1$ acting on a Hilbert space. We use the ket notation (\emph{e.g.}~$\ket{\Psi}$)
to denote pure quantum states. Quantum measurements
are described by Positive Operator Valued Measures (POVMs) which are finite sets of positive
operators that sum up to the identity, \emph{e.g.}~$\{P_x, x\in \mathcal{X} : P_x\geq 0\ \&\ \sum_x P_x = \id \}$,
where $\mathcal{X}$ is a finite set of indices. If for some measurement the operators $P_x$ are mutually orthogonal
projectors, then we say that the measurement is projective. The probability of observing outcome $x\in\mathcal{X}$
when measuring the state $\rho$ with the measurement described by $\{P_x\}_{x \in \mathcal{X}}$ is given by
$p_x:=\tr(P_x \rho)$.}
{W}e use $X_j^n$ as a shorthand for the string $X_j,\ldots,X_n$ ($j\leq n$).
The symbol $\approx_{\epsilon}$ will be used to express that two states are $\epsilon$-close in the trace distance
(\emph{e.g.}~$\sigma \approx_{\epsilon} \rho$ for two states $\rho,\sigma$). In several occasions
we will denote $\mathcal{R}$ for a family (not necessarily specified) of $2$-universal hash functions.
We will use $X\in_{R} \mathcal{E}$ to say that $X$ is picked uniformly at random from set $\mathcal{E}$.
$[n]$ is a shorthand notation for $\{1,\ldots,n\}$.

\section{Results}


In this section, we will present the results of our work. Formal statements and their proofs
will be given in the Methods Section.

\begin{itemize}
  \item We start by presenting the MDI protocols for OT and BC for the case where
  the honest parties have access to perfect single photon sources.
  \item Then we present and analyze the security of a protocol for BC where the honest parties only have imperfect single photon
  sources, \textit{i.e.}~multiphoton emissions are possible.
  \item Finally we show that there is a family of protocols that
  cannot be secure for OT in MDI settings when the honest parties are using imperfect
  single photon sources.

\end{itemize}

\subsection{Bit Commitment (BC) with perfect single photon sources}
\label{Sec:BC_perf}

In this section we explain Bit Commitment, present a protocol that implements it
when the honest parties have access to a perfect single photon source (Protocol \ref{Ptol:BC_perfect_source}),
and  state the security of this protocol in the Noisy Quantum Storage Model.

Bit Commitment is a two-phase task between two parties, Alice and Bob, where in the
first phase Alice commits to a bit of her choice to Bob. Later they can run the second phase (the ``Open'' phase)
 where Alice reveals the bit to which she committed. Importantly, Alice should not be able
to open a bit different than the one to which she committed. Also we require that Bob
cannot learn the value of the committed bit before Alice opens it. The case in which
 Alice commits to a bit-string rather than a single bit is called String
 Commitment. In the
following we give a formal definition for a randomized version of String Commitment,
where Alice does not get to choose the string she commits to. This string will be produced
uniformly at random by the protocol. Note that a Randomized String Commitment can be
turned into a String Commitment scheme as explained in \cite{Steph_1}.

In this paper we will use the security definition of Bit Commitment from \cite{Steph_1} informally stated below.
The reader can find the formal Definition \ref{Def:Sec_ideal_BC} in Appendix \ref{Sec:Definitions}.

\begin{Def}[Randomized String Commitment (informal)]\label{Def:BC_informal}\hfill \\
  A protocol implements an $(l,\epsilon)$-Randomized String Commitment if it satisfies the following three conditions:
    \begin{description}
    \item[Correctness] If both Alice and Bob are honest, the protocol outputs a classical state $\rho_{C_1^l C_1^l F}$ such that
    $\rho_{C_1^l F}$ is  $\epsilon$-close to
    $\tau_{C_1^l} \otimes \ketbra{accept}{accept}_F$, where $\tau_{C_1^l}:=\frac{\id}{2^l}$ is maximally mixed and $C_1^l$ is an $l$-bit-string.
    \item[Security for Bob] If Bob is honest, {then} there exists a string $C_1^l$ after the Commit phase, such that the probability
    that Alice opens to another string ${C_1^l}'\neq C_1^l$, and Bob accepts is smaller than $\epsilon$.
    \item[Security for Alice] If Alice is honest, then after the Commit phase and
    before the Open phase Bob is ``$\epsilon$-ignorant'' about the string $C_1^l$ that Alice has received during the Commit phase.
  \end{description}
\end{Def}

\begin{Myfigure}
  \begin{tikzpicture}[scale=1,line width=1.2pt]
    \node[draw, rounded corners=2pt] (Bob) at (4,0){Bob};
    \node[draw, rounded corners=2pt] (Alice) at (-4,0){Alice};
    \node[draw, rounded corners=2pt] (Commit) at (0,0){Commit};
    \draw[->] (Commit) -- (Alice)node[pos=0.5,above]{$C_1^l\in_R\{0,1\}^l$};
    \draw[->] (Commit) -- (Bob)node[pos=0.5,above]{``Committed''};

    \draw[-,dashed] (-1,-1) -- (1,-1);

    \node[draw, rounded corners=2pt] (Bob2) at (4,-2){Bob};
    \node[draw, rounded corners=2pt] (Alice2) at (-4,-2){Alice};
    \node[draw, rounded corners=2pt] (Open) at (0,-2){Open};
    \draw[<-] (Open) -- (Alice2)node[pos=0.5,above]{``Open''};
    \draw[->] (Open) -- (Bob2)node[pos=0.5,above]{$C_1^l$};

  \end{tikzpicture}
  
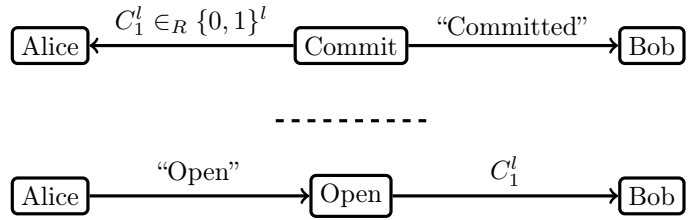
\captionof{figure}{Ideal Randomized String Commitment.
  In the first part Alice gets a random $l$-bit string $C_1^l\in\{0,1\}^l$, and Bob is notified that
  the string is committed. In the second phase, Alice asks ``the box'' to reveal the string to Bob.}
\end{Myfigure}

In this work we show that the protocol below implements a secure String Commitment scheme.

\begin{Thm*}[Security of Protocol \ref{Ptol:BC_perfect_source} (Informal)]
  {Let $0<\epsilon<1$, let $l$ be the length of the string to be committed, let
  $e_{\rm err}$ be the expected error rate between the outcomes of honest Alice and honest Bob in the
  preparation phase of the protocol \ref{Ptol:BC_perfect_source}, and let $D$ be an upper-bound on the size of
  dishonest Bob's quantum memory expressed in qubits.}
  If honest players have access to perfect single photon sources, then Protocol \ref{Ptol:BC_perfect_source}
  implements an $(l,3 \epsilon)-$Randomized String Commitment
  according to the above definition. In particular
  it does so using $n$ rounds {of quantum communication}, where $n$ is a positive integer solution to
  $({\lambda-h(\delta)}) n \geq {l+2\log(1/2\epsilon)+\ln(\epsilon^{-1})}$,
  where $\lambda:=f(-D/n)-1/n$ ($f$ is defined in eq.~\eqref{eq:def_f1_1}), and $\delta=2 e_{\rm err} + 2 \alpha_2$, where
  $\alpha_2$ is a term that accounts for statistical fluctuations $\alpha_2=\mathcal{O}(n^{-1/2})$.
\end{Thm*}

\noindent The reader can find a formal version of this theorem in the Methods Section together with
its proof, see Theorem \ref{Thm:Sec_BC_perfect}. Intuitively -- in the MDI settings with perfect single photon sources
-- the only difference
for the security analysis as compare to
the analysis of the protocols presented in Refs.~\cite{Steph_1,NJCKW12} is that honest Bob sends information to malicious Alice. However
since we are guarantied (by assumption) that Bob sends BB84 states on single photons,
we can use a purification argument in order to reduce the MDI situation to the one of Refs.~\cite{Steph_1,NJCKW12}
(see Figure \ref{fig:dishonest_Alice_equivalence}) where only Alice sends information to Bob. \newline

{\begin{Rmk}[Bell measurement]\label{Rmk:Bell_meas}
  In the MDI protocols we will describe in this paper we use measurements that we call ``Bell measurement''.
  Usually the terminology ``Bell measurement'' designates a two-qubits-projective measurement described by
  the four projections onto the Bell states $X^aZ^b \tfrac{\ket{00}+\ket{11}}{\sqrt{2}},\ (a,b)\in\{0,1\}^2$,
  where $X$ and $Z$ denote the Pauli $X$ and $Z$ operators. In general, a measurement whose operators are projections onto
  four orthogonal maximally entangled state is called a ``deterministic Bell measurement''. However, in this
  paper the expression ``Bell measurement'' refers to a more general type of measurements sometimes called
  ``probabilistic Bell measurements''. A probabilistic Bell measurement is a two-qubit measurement where one or two
  of its operators are projections onto orthogonal maximally entangled states, the other operators being arbitrary
  (on the condition that the set of operators considered describes a valid measurement).
  The outcomes corresponding to operators that are not projections onto maximally entanglement states will be considered as
  `failure'' outcomes. This notion of probabilistic Bell measurement arises naturally when
  considering linear optical implementation of such measurements. Indeed linear optics does not
  allow to implement deterministic Bell measurements \cite{CL01}.
  \newline
  Furthermore, it is sometimes possible to detect when the qubits were lost before
  reaching the measurement device. This will also be considered as a ``failure'' outcome.
  The overall probability of obtaining a failure outcome is denoted $p_{\rm fail}$.
\end{Rmk}}

We present below a protocol for Randomized String Commitment adapted from \cite{Steph_1} to
the measurement-device-independent case. In this protocol
Alice and Bob will start with a preparation phase in which they send $n$ states randomly chosen from the
set $\{\ket{0},\ket{1},\ket{+},\ket{-}\}$ to the measurement station which will perform
a Bell measurement on these qubits and broadcast the outcome. For the rounds in which Alice and
Bob have used the same basis to encode their states, the Bell measurement outcome tells
Bob whether he has encoded the same bit as Alice in his qubit or the opposite bit. If they have used a different
basis then the Bell measurement outcome does not give any information on their correlation. In order to
force any dishonest party to store quantum information, both parties will wait a certain time $\Delta t$
before Alice reveals to Bob which bases she has used to prepare her qubits. This allows Bob
to compute the set of rounds $\mathcal{I}\subseteq [n]$ where they have used the same bases. Bob
will discard the rounds that do not belong to $\mathcal{I}$. From there, they will
only use classical communication to extract a random committed string $C_1^l$ in the Commit phase, and to
reveal this string in the Open phase.

For the following protocol, we will use a randomly generated $[n,k,d]$-linear code
$\mathcal{C} \subseteq \{0,1\}^n$ with fixed rate $R:=k/n$ to describe
Protocol \ref{Ptol:BC_perfect_source} and to analyze its security. This does not affect the efficiency of the
protocol since the honest parties do not need to decode: We only need to use this code
to impose that two strings with the same syndrome have Hamming distance at least $d$. We denote ${ \textsf{Syn}: \{0,1\}^n \mapsto \{0,1\}^{n-k}}$ for
the function that outputs the parity-check syndrome of the code $\mathcal{C}$. In this protocol
we use the two following shorthand notations
$\alpha_1:=\sqrt{\frac{\ln \epsilon^{-1}}{2 n}}$, $\alpha_2:=\sqrt{\frac{\ln \epsilon^{-1}}{2(1/2-\alpha_1)n}}$. Let
$f(\cdot)$ be the function defined as follows.
\begin{align}\label{eq:def_f1_1}
  f(x):=
  \begin{cases}
    {0} &\text{ {if} }\ {x<-1}\\
    g^{-1}(x) &\text{ if }\ -1\leq  x <1/2\\
    x &\text{ if }\ {1/2\leq x  \leq 1},
  \end{cases}
\end{align}
where $g(x):= h(x)+x-1$ and $h(x):= -x \log(x)-(1-x)\log(1-x)$ is the binary entropy.

\begin{framed}
\begin{Ptol}[Randomized String Commitment]\label{Ptol:BC_perfect_source}
  \hfill \\
  \textbf{Inputs:} security parameter $\epsilon>0$, length of the committed string $l>0$,
  bound on the size of the adversary's quantum memory $D$, $e_{\rm err}$ is the expected error rate
  that should be observed between Alice's an Bob strings $X_{\mathcal{I}}$ and $\hat X_{\mathcal{I}}$ (see below).

  \begin{description}
    \item[Preparation phase]\hfill\\
    Choose the number $n$ of rounds that click, such that $n\geq \frac{l+2\log(1/2\epsilon)+\ln(\epsilon^{-1})}{\lambda-h(\delta)}$, where
    $\lambda:=f(-D/n)-1/n$, and $\delta=2 e_{\rm err} + 2 \alpha_2$.
  \begin{enumerate}
  \item For round $i$ (until the number of rounds in which the measurement station has clicked is higher than $n$):
    \begin{itemize}
      \item Alice chooses $X_i\in_R \{0,1\}$ and $\Theta_i \in_R \{0,1\}$ uniformly at random,
    and prepares and sends the state $\ket{X_i}_{\Theta_i}$
    (where $\ket{0}_0:=\ket{0}, \ket{1}_0:=\ket{1},\ket{0}_1:=\ket{+},\ket{1}_1:=\ket{-}$) to the measurement station.
      \item Bob chooses $\hat X_i \in_R \{0,1\}$ and $\hat \Theta_i \in_R \{0,1\}$ uniformly at random
    and prepares and sends the state $\ket{\hat X_i}_{\hat \Theta_i}$
    (where $\ket{0}_0:=\ket{0}, \ket{1}_0:=\ket{1},\ket{0}_1:=\ket{+},\ket{1}_1:=\ket{-}$) to the measurement station.
      \item The measurement station performs a Bell measurement on the two states it receives, and broadcasts the
      outcome, or whether the measurement failed {(see Remark \ref{Rmk:Bell_meas})}. Depending on the outcome,
      Bob chooses whether he should flip his bit or not.
    \end{itemize}
  \item Alice and Bob discard all the rounds where a failure has been announced. Let's call $n$ the remaining number
  of rounds. Alice has strings $X_1^n$ and $\Theta_1^n$ $\in \{0,1\}^n$, and Bob has strings
  $\hat X_1^n$ and $\hat \Theta_1^n$ $\in \{0,1\}^n$.
  \item Both parties wait for a time $\Delta t$.
  \item Alice sends $\Theta_1^n$ over to Bob.
  \item Bob computes the set $\mathcal{I} \subseteq [n]$ of rounds $i$ where $\Theta_i= \hat \Theta_i$.
  Bob discards all the rounds $j \notin \mathcal{I}$. Let's then
  call $\hat X_\mathcal{I}$ the string formed by all the remaining bits $\hat X_i$ with $i\in \mathcal{I}$.

  Note that when there
  is no noise we have that $\forall i \in \mathcal{I}$ $X_i=\hat X_i$. In practice there are always errors: We will
  call $e_{\rm err}$ the expected error rate between $X_i$ and $\hat X_i$ (for $i\in \mathcal{I}$), in other words
   $e_{\rm err}$ is the expected fraction of error between $X_{\mathcal{I}}$ and $\hat X_{\mathcal{I}}$.

  \end{enumerate}
  \item[Commit Phase]\hfill\\
  \begin{enumerate}
    \item Bob checks whether $m:=|\mathcal{I}| \geq 1/2 \cdot n - \alpha_1$. If it is not the case Bob
    aborts the protocol.
    \item Alice chooses a random $[n,k,d]$-linear code $\mathcal{C}$ (for fixed $n$ and $k$) and computes $w=\textsf{Syn}(X_1^n)$ and sends it to Bob.
    \item Alice picks a random 2-universal hash function $r \in_R \mathcal{R}$ and sends it to Bob.
    \item Alice outputs $C_1^l:={\rm Ext}(X_1^n,r)$ where ${\rm Ext(\cdot,\cdot)}$ is a randomness extractor
    from the 2-universal family of functions.
  \end{enumerate}

  \item[Open phase]\hfill\\
  \begin{enumerate}
    \item Alice sends $X_1^n$ to Bob.
    \item Bob computes its syndrome and checks if it agrees with $w$ he received from Alice in the Commit phase. If they disagree
    Bob aborts the protocol.
    \item Bob checks that the number of rounds $i\in \mathcal{I}$ where $X_1^n$ and $\hat X_\mathcal{I}$ do not agree lies in
    the interval $]e_{\rm err}- \alpha_2, e_{\rm err} +\alpha_2[$. If not, Bob aborts the protocol, otherwise
    Bob accepts, and he outputs $C_1^l:={\rm Ext}(X_1^n,r)$ where ${\rm Ext(\cdot,\cdot)}$ is a randomness extractor
    from the 2-universal family of function.
  \end{enumerate}
\end{description}
In order to satisfy the security definition for Randomized String Commitment, when the protocol aborts, the honest parties will continue
the protocol as if they were not aborting -- in particular they do not announce the abort event until the end of the protocol -- and in the end
honest Bob always rejects the commitment and output a uniformly random value to $\tilde C_1^l$, and honest Alice
outputs a uniformly random value for $C_1^l$.
\end{Ptol}
\end{framed}

\subsection{Oblivious Transfer (OT) with perfect single photon sources}\label{Sec:OT_perf}

In this section we explain OT, present a protocol that implements it
when the honest parties have access to a perfect single photon source (Protocol \ref{Ptol:OT}),
and we state the security of this protocol in the Bounded Quantum Storage Model.

\begin{Myfigure}\label{Fig:OT}
  \begin{tikzpicture}[scale=1,line width=1.2pt]
    \node[draw, rounded corners=2pt] (Bob) at (4,0){Bob};
    \node[draw, rounded corners=2pt] (Alice) at (-4,0){Alice};
    \node[draw, rounded corners=2pt, minimum width=4em] (OT) at (0,0){OT};
    \draw[->] (Commit) -- (Alice)node[pos=0.5,above]{$(S_0,S_1)$};
    \draw[->] (Commit) -- (Bob)node[pos=0.5,above]{$(S_C,C)$};
  \end{tikzpicture}
  
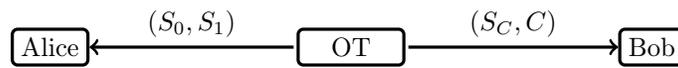
\captionof{figure}{In a Randomized $1$-out-$2$ Oblivious String Transfer, Alice should get two random $l$-bit strings
  $(S_0,S_1)$ and Bob should receive a random bit $C$ together with $S_C$ which is one
  of the two strings Alice has received. Alice should never learn $C$ and Bob should remain ignorant about at least one
  of the two bit-strings Alice receives.}
  \label{Fig:OT}
\end{Myfigure}

OT, or rather its variant called Randomized 1-out-2 Oblivious String Transfer, is a task where Alice receives two
random strings $(S_0,S_1)$, and Bob receives one of this string $S_C$ together with its corresponding index $C$ (see Fig.~\ref{Fig:OT}).
We will use the definition of the Randomized 1-out-2 Oblivious String Transfer from \cite{Steph_1} which is informally stated
below. The reader can find the formal Definition \ref{Def:Sec_ideal_OT} in Appendix \ref{Sec:Definitions}
\begin{Def}[Randomized String Transfer (informal)]\label{Def:OT_informal}
  A protocol implements an $(l,\epsilon)$-Randomized 1-out-2 Oblivious String Transfer if it satisfies the following three
  conditions:
  \begin{description}
    \item[Correctness] If Alice and Bob are honest the protocol's output state $\rho_{(S_0,S_1),(S_C,C)}$ is such that
    the reduced state $\rho_{S_0,S_1,C}$ is $\epsilon$-close to $\tau_{S_0}\otimes \tau_{S_1} \otimes \tau_{C}$, where
    $\tau_R$ denotes the maximally mixed state on register $R$, and $S_0, S_1$ are two $l$-bit-strings.
    \item[Security for Alice] If Alice is honest, then Alice should get two $l$-bit-strings $S_0$ and $S_1$ such that
    there exists a binary random variable $\tilde C$
    such that Bob is ``$\epsilon$-ignorant'' about the bit string $S_{1-\tilde C}$. We say that the protocol is $\epsilon-$hiding.
    \item[Security for Bob] If Bob is honest then he should receive a random bit $C$ and an
    $l$-bit-string $\hat S_{C}$, such that Alice is ``$\epsilon$-ignorant'' about $C$. We say that the protocol is $\epsilon-$binding.
  \end{description}
\end{Def}

In this work we show that Protocol \ref{Ptol:OT} presented below implements a secure
Randomized Oblivious Transfer.

\begin{Thm*}[Randomized 1-out-2 OT (Informal)]
  {Let $0<\epsilon<1$, let $l>0$ be an integer, let $e_{\rm err} \in ]0,1/2[$ be the
  expected error rate between the outcomes of honest Alice and honest Bob in the preparation phase of Protocol \ref{Ptol:OT}, and
  let $D$ be an upper-bound on the size of dishonest Bob's quantum memory expressed in qubits.}
  When honest parties have access to perfect single photon sources, Protocol \ref{Ptol:OT} implements
  an $(l,\epsilon)$-Randomized 1-out-2 Oblivious String Transfer according to the above definition in the Bounded Quantum Storage
  Model.
  In particular
  it does so using a linear (in the length $l$ of Alice's strings $|S_0|=|S_1|=l$) number of rounds of quantum communication.
  More precisely, the number $n$ of quantum communication rounds must satisfy $n\geq 2\, \frac{l+D+1-2\log(1-\sqrt{1-\epsilon^2})}{\lambda - h(e_{\rm err}) -\mathcal{O}(n^{-1/2})}$,
  where $\lambda:=1/2-\delta'$ with $\delta'=(2-\log(\sqrt{(32 \ln \epsilon^{-1})/n}))\sqrt{(32 \ln \epsilon^{-1})/n}$
\end{Thm*}

The reader can find a formal version of this theorem in the Methods Section together with
its proof, see Theorem \ref{Thm:Sec_OT_perf}.
Intuitively -- in the MDI settings with perfect single photon source
--  using a purification argument on the states sent by Bob, we can essentially
reduce the security proof of our protocol to the security proofs of the trusted device protocol presented in Ref.~\cite{Steph_1}
in which all devices are trusted. However we need to be careful because
we also want to take into account noise which has not been done in Ref.~\cite{Steph_1}.

The protocol presented below is also adapted from \cite{Steph_1}.
For the following Protocol, let $\alpha_1:=\sqrt{\frac{\ln \epsilon^{-1}}{2 n}}$ be a term accounting for
statistical fluctuations.

\begin{framed}
\begin{Ptol}[Randomized 1-out-2 OT] \label{Ptol:OT}\hfill \\
  \textbf{Inputs:} security parameter $\epsilon>0$, the length $l$ of the strings Alice receives, the bound (expressed in qubits) on the adversaries memory $D$,
  expected error rate $e_{\rm err}$ between Alice's and Bob's strings $X_{\mathcal{I}}$ and $\hat X_{\mathcal{I}}$ defined below.

  \begin{description}
    \item[Preparation phase]
      They first choose the number of rounds $n$ in which the station clicks, such that $n\geq 2\, \frac{l+D+1-2\log(1-\sqrt{1-\epsilon^2})}{\lambda - h(e_{\rm err}) -\mathcal{O}(n^{-1/2})}$.
      Then Alice and Bob do the same as in the preparation phase of Protocol \ref{Ptol:BC_perfect_source}. At this point
      Alice has a string $X_1^n$, and Bob has a string $\hat X_\mathcal{I}$ and the set $\mathcal{I} \subseteq [n]$.
    \item[Post Processing] \hfill \\
    \begin{enumerate}
      \item Bob checks whether $|\mathcal{I}| \geq (1/2-\alpha_1) n =:m$. If this is the
      case he randomly truncates $\mathcal{I}$ such that $|\mathcal{I}|=m$. Otherwise he aborts.
      \item Bob picks a random subset of $\mathcal{I}^c$ of size $m$ called $\mathcal{I}_{\rm Bad}$.
      Bob chooses a bit $C$ uniformly at random. He then renames $(\mathcal{I},\mathcal{I}_{\rm Bad})$
      into $(I_C,I_{1-C})$. Bob sends $(I_0,I_1)$ to Alice.
      \item Alice sends Bob error correction information for the strings $X_{I_0}$ and $X_{I_1}$.
      \item Bob uses the error correction information $O$ to correct his string $\hat X_\mathcal{I}$.
      \item Alice chooses two $2$-universal hash functions $r_0,r_1 \in_R \mathcal{R}$ uniformly at random and sends them
      to Bob.
      \item Alice outputs $(S_0,S_1):=\big({\rm Ext}(X_{I_0},r_0), {\rm Ext}(X_{I_1},r_1)\big)$, and Bob
      outputs $(\hat S_C,C):=\big({\rm Ext}(X_{I_C},r_C), C\big)$.
    \end{enumerate}
  \end{description}
  In order to satisfy the security definition for OT, when an honest party aborts the protocol, the aborting party will continue
  the protocol as if they were not aborting -- in particular, they do not announce the abort event until the end of the protocol --
  except that in the end, when the abort event is announced all honest parties assign to their outputs
  uniformly random values.
\end{Ptol}
\end{framed}

\subsection{Bit Commitment with imperfect single photon sources}

In this section we present a protocol that implements String Commitment
when the honest parties \emph{do not} have access to perfect single photon sources (Protocol \ref{Ptol:BC_imperfect_source}),
and we state the security of this protocol in the Noisy Quantum Storage Model.
In this situation, the multiphoton emissions can leak -- to dishonest Alice -- information about the bases
Bob used in his encoding. As a consequence, malicious Alice could take advantage of that
by selectively announcing all single photon emissions as ``lost'', and keep only the
rounds where she has information on the bases used by Bob. Malicious Bob can do the same
to get some advantage over honest Alice. To prevent this, and make sure that most of the
rounds that are kept in the end correspond to single photon emission rounds we will
use the decoy states technique \cite{LMC05} similar to \cite{WSEE}. This will allow the honest party to estimate an upper-bound
on the number of rounds that are kept in the end and which correspond to
multiphoton emissions.\\

Examples of photon sources are lasers. They produce coherent states that can be written
in the Fock basis as follows:
\begin{align}
  \ket{\alpha}=e^{-\frac{|\alpha|^2}{2}}\sum_{n=0}^{\infty} \frac{\alpha^n}{\sqrt{n!}} \ket{n},
\end{align}
where $\ket{n}$ is the photon number eigenstate associated to photon number $n$, and $\alpha \in \mathbb{C}$.
The intensity is the average number of photons of such a state, and is given by $|\alpha|^2$. As in some MDI QKD experiments \cite{LCQ12,LCW13,TLX14}
one can use a randomized phase coherent state in order to turn the laser into an imperfect single photon source {\cite{WSEE}}. A randomized
phase coherent state is a coherent state where $\alpha=r e^{i\phi}$ with $r>0$ and
where $\phi$ is chosen uniformly at random in $[0,2\pi[$. To anyone that does not know which phase has been picked, this state
is equivalent to the mixed state $\rho_{|\alpha|^2}=\sum_{n=0}^\infty e^{-|\alpha|^2} \frac{{|\alpha|^2}^n}{n!} \ketbra{n}{n}$.
When one wants to produce single photons, one can use an attenuated laser that produces states with
a low average number of photon, \textit{i.e.}~with small $|\alpha|^2$. For example for
$|\alpha|^2=0.1$, the state $\rho_{|\alpha|^2}$ is essentially a mixture of $\ketbra{0}{0}$ with probability
$\approx 0.905$, $\ketbra{1}{1}$ with probability $p_1\approx 0.0905$, and multiphoton emissions with probability
$p_{\geq 2} \approx 0.0045$, which gives a fraction of $\approx 5\%$ of multiphoton emissions conditioned on
emitting at least one photon, which means that the source mostly (95\% of non $0$ emissions) emits single photons and
and emits a small amount of multiphoton states (about $5\%$ of non $0$ photon emissions). In a protocol like MDI BC
we encode the state in some degree of freedom like polarization. This is a problem for the rounds
where multiple photons have been emitted. When only one photon is emitted the possible states
Bob can encode are $\{\ket{0},\ket{1},\ket{+},\ket{-}\}$, and therefore the state sent from Bob to Alice
conditioned on a choice of basis, $\theta=0$ or $\theta=1$ are
$\rho_{|\theta=0}=1/2(\ketbra{0}{0}+\ketbra{1}{1})=\id/2=1/2(\ketbra{+}{+}+\ketbra{-}{-})=\rho_{|\theta=1}$, meaning that Alice cannot guess
which basis Bob has used to encode his state. On the contrary, if for example two photons have been emitted
the states are $\rho_{|\theta=0}=1/2(\ketbra{00}{00}+\ketbra{11}{11}) \neq 1/2(\ketbra{++}{++}+\ketbra{--}{--})=\rho_{|\theta=1}$
meaning that Alice can guess the basis used with non $0$ advantage. This is a problem since
security against dishonest Alice {relies} on her being ignorant about Bob's basis information. In particular we
want to avoid the case where dishonest Alice measures the photon number of the incoming state from Bob,
and chooses to announce failure only if she receives single photon. This is why we use decoy states: They
will allow us to estimate how many single photon rounds have been reported as failure.\\

For BC in the case where honest parties use imperfect single photon sources,
Protocol \ref{Ptol:BC_imperfect_source} can be used. The main difference as compare to Protocol
\ref{Ptol:BC_perfect_source} is the use of $q$ additional decoy states in the ``Preparation phase''. Alice (Bob)
can use different intensities\footnote{We remind the reader that intensities correspond to the mean number of photons
produced by the source. For a (randomized phase) coherent state the intensity is given by $|\alpha|^2$. In many practical cases
the intensity of the source can be chosen.} for the state she (he) sends. Among these intensities one
will correspond to the ``signal'' state and will be denoted $a_s$ ($b_s$), while the others
will be the ``decoy'' states with intensities $a \in \{a_{d_1} \ldots a_{d_q}\}$ ($b \in \{b_{d_1} \ldots b_{d_q}\}$).
In Protocol \ref{Ptol:BC_imperfect_source} we will call $n_1^A+n_{\geq2}^A$ ($n_1^B+n_{\geq2}^B$) the
number of rounds where Alice (Bob) has used a ``signal'' state -- \textit{i.e.}~a state with intensity
$a_s$ ($b_s$) -- and where the measurement station reported the measurement as successful. $n_1^A$ ($n_1^B$) is
the number of these states where Alice (Bob) has sent $1$ photon, and $n_{\geq 2}^A$ ($n_{\geq 2}^B$)
is the number of these rounds where Alice (Bob) has sent $\geq 2$ photons. Note that at
the end of Step 1 of the ``preparation phase'', and because we do \emph{not} consider dark counts in this work, Alice (Bob) knows the value
of $n_1^A+n_{\geq2}^A$ ($n_1^B+n_{\geq2}^B$). However even if she (he) knows the sum $n_1^A+n_{\geq2}^A$ ($n_1^B+n_{\geq2}^B$),
she (he) does \emph{not} know the individual terms $n_1^A$ ($n_1^A$) and $n_{\geq2}^A$ ($n_{\geq2}^B$) of this sum. Alice (Bob) will only be able to
estimate a lower-bound $L_{A1}$ ($L_{B1}$) on $n_1^A$ ($n_1^B$) by using the decoy states. Since $n_1^A+n_{\geq2}^A$ ($n_1^B+n_{\geq2}^B$)
is known to Alice (Bob), this lower-bound gives automatically an upper-bound $U_{A2}=n_1^A+n_{\geq2}^A - L_{A1}$
($U_{B2}=n_1^B+n_{\geq2}^B - L_{B1}$) on $n_{\geq 2}^A$ ($n_{\geq 2}^B$).\\

In the following we will write $p_a$ ($p_b$) for the probability that Alice (Bob) prepares a signal of
intensity $a \in \{a_s,a_{d_1} \ldots a_{d_q}\}$ ($b \in \{b_{a_s},b_{d_1} \ldots b_{d_q}\}$). When the identity
of the emitter is not determined, the intensity will be denoted $i$ (meaning that $i=a$ is the emitter is Alice
or $i=b$ is the emitter is Bob).
The probability that an emitter emits $k$ photons will be denoted $p_k$ (\emph{e.g.}~if $k=1$ then we will write
$p_1$ etc.). The probability that the emitter emits more than $k$ photons will be denoted
$p_{\geq k}$ (\emph{e.g.} $p_{\geq 2}$). We will also mix the two above notations when
talking about conditional events. For example, the probability that Alice emits $2$ photons
conditioned on choosing signal intensity $a$ will be denoted $p_{2|a}$.\\

In this paper we show that Protocol \ref{Ptol:BC_imperfect_source} below is secure,
in particular we show the following. Here we state this lemma in the case in which Alice
is honest. A similar statement would apply for honest Bob.

\begin{Lmm*}[{Single}-photon emission round number estimation (Informal)]
  {Let $q\geq 2$ be the number of different intensities honest Alice can use for the decoy states
  in Protocol \ref{Ptol:BC_imperfect_source}. Let $\{ p_{a_{d_1}} , \ldots, p_{a_{d_q}} \}$ be the (known) probabilities
  that Alice's source emits a state with (known) intensities $\{ a_{d_1}, ..., a_{d_q}\}$. Let
  $p_s$ be the (known) probability that Alice's source emits a state with (known) intensity $a_s$ corresponding to the signal
  state (\emph{i.e.}~non-decoy state). Let $x^i$ be the observed number of non-discarded rounds where
  Alice has prepared a signal of intensity $i \in \{a_s, a_{d_1}, ..., a_{d_q}\}$. Then with high probability,
  the number of non-discarded signal rounds $n^A_1$ in which Alice's source has emitted exactly $1$ photon is lower-bounded
  by $L_{A1}$, where $L_{A1}$ is a function of the intensities $\{a_s, a_{d_1}, ..., a_{d_q}\}$, the probabilities
  $\{ p_s,p_{a_{d_1}} , \ldots, p_{a_{d_q}} \}$, and the observations $\{x^{a_s},x^{a_{d_1}},\ldots,x^{a_{d_q}} \}$. The analytical expression for $L_{A1}$ is given in the
  formal version of the lemma, Lemma \ref{Thm:estimation} in the case $q=2$. Its proof is
  given in Appendix \ref{Sec:Thm:estimation}. For $q>2$ one can compute $L_{A1}$
  numerically as explained in Appendix \ref{Sec:Thm:estimation}.
  }
\end{Lmm*}

The above lemma is an essential ingredient to prove the following security theorem.

\begin{Thm*}[Security of Protocol \ref{Ptol:BC_imperfect_source} (Informal)]
  {Let $0<\epsilon<1$, let $l$ be the length of the committed string, let $p$ be the
  probability that -- in the preparation phase of an honest execution of Protocol \ref{Ptol:BC_imperfect_source} --
  a given round $i$ is not discarded, let $e_{\rm err}$ be be the expected error rate between
  the outcomes of honest Alice and honest Bob in the Preparation phase of the protocol, and let $D$ be a bound on the size of dishonest Bob's
  quantum memory measured in qubits. }
  Protocol \ref{Ptol:BC_imperfect_source} implements an $(l,\epsilon)-$Randomized String Commitment as
  defined in Definition \ref{Def:BC_informal}. In particular it does so using a number $N$ of
  quantum communication rounds that is linear in $l$. More precisely $N$ must
  satisfy $(p-\sqrt{\ln(\epsilon^{-1})/2N})\, N\geq n^\star$ where
  $n^\star$ is smallest {positive} integer solution to
  $(\lambda-h(\delta))n \geq {l+2\log(1/2\epsilon)+\ln(\epsilon^{-1})}$,
  $\lambda:= f(-D/n) -(\gamma+\alpha_4^A)-1/n$ with $n$ being the length of honest Alice's string $X_1^n$
  produced at the end of the Preparation phase, $\alpha_4^A$ is a term accounting for statistical fluctuations, and $\delta$ is a function of the expected error rate $e_{\rm err}$.
  The exact expression of $\delta$ and $\alpha_4^A$ are given in
  the formal version of this theorem: Theorem \ref{Thm:security_BC_imperfect}.
\end{Thm*}

A formal version of this theorem together with its proof are given in the
Methods Section: Theorem \ref{Thm:security_BC_imperfect}.

In Protocol \ref{Ptol:BC_imperfect_source} and its security analysis we will use the
following notations: $\epsilon\in ]0,1[$, and $f_{a_s}, f_{b_s} \in [0,1]$ are fractions defined in Step 2 of Protocol \ref{Ptol:BC_imperfect_source}.
$\alpha_1, \alpha_2$ are the same as in Section \ref{Sec:BC_perf}, and as for the terms
$\beta^A, \beta^B,\alpha_4^A,\alpha_4^B$, they account for statistical fluctuation. They all are $\mathcal{O}(1/\sqrt{N})$ where $N$
is the number of rounds of the protocol. Their exact expressions are given
in Theorem \ref{Thm:security_BC_imperfect}. As in Protocol \ref{Ptol:BC_perfect_source},
$\mathcal{C}$ is an random $[n,k,d]$-linear code and ${\rm Syn}:\{0,1\}^n \mapsto \{0,1\}^{n-k}$ is the function that
outputs the parity-check syndrome of code $\mathcal{C}$.

In order to satisfy the security definition for Randomized String Commitment (Def.~\ref{Def:BC_informal}), when the protocol aborts, the honest parties will continue
the protocol as if they were not aborting -- in particular they do not announce the abort event until the end of the protocol -- and in the end
honest Bob always rejects the commitment and assigns a uniformly random value to his output $\tilde C_1^l$, and honest Alice
assigns a uniformly random value to her output $C_1^l$.

\begin{framed}
\begin{Ptol}[Randomized String Commitment with decoy states]\label{Ptol:BC_imperfect_source}
  \hfill \\
  \textbf{Inputs:} The security parameter $\epsilon>0$, the parameter $\gamma \in[0,1/2]$
  that essentially measures how good the single photon sources are,
   the length $l$ of the string that will be produced by the protocol, the maximum
  size (expressed in qubits) of the adversary's quantum memory $D$, the expected
  error rate $e_{\rm err}$ between Alice's and Bob's string $X_\mathcal{I}$ and
  $\hat X_{\mathcal{I}}$, the probability distributions $(p_{a_s},p_{a_{d_1}},\ldots, p_{a_{d_q}})$ and $(p_{b_s},p_{b_1},\ldots, p_{b_q})$ that Alice and Bob use
  intensities $\{a_s,a_{d_1},\ldots,a_{d_q}\}$ and $\{b_s,b_{d_1},\ldots,b_{d_q}\}$ respectively.
  \begin{description}
    \item[Preparation phase]\hfill\\
    Alice and Bob agree on a number $N$ of rounds. $N$ must satisfy $(p-\sqrt{\ln(\epsilon^{-1})/2N})\, N\geq n^*$,
     where $n^*$ the smallest positive integer solution to the inequality eq.~\eqref{eq:cond_thm}, and
    where $p$ is the probability that any given round $i\in [N]$ is not discarded in the preparation phase when both parties are honest.
    \vspace{-1em}
  \begin{enumerate}
  \item For round $i\in [N]$:
    \begin{itemize}
      \item Alice chooses $X_i\in_R \{0,1\}$ and $\Theta_i \in_R \{0,1\}$ uniformly at random, and chooses intensity
      $a\in \{a_s, a_{d_1}\ldots a_{d_q}\}$ with some probability distribution $p_a$. Alice prepares a quantum
      signal of intensity $a$, encoding $X_i$ in the basis $\Theta_i$, and sends it over to the measurement station.
      \item Bob chooses $\hat X_i\in_R \{0,1\}$ and $\hat \Theta_i \in_R \{0,1\}$ uniformly at random, and chooses intensity
      $b\in \{b_s, b_{d_1}\ldots b_{d_q}\}$ with some probability distribution $p_b$. Bob prepares a quantum
      signal of intensity $b$, encoding $\hat X_i$ in the basis $\hat \Theta_i$, and sends it over to the measurement station.
      \item The measurement station performs a Bell measurement on the two states it receives, and publicly reveals the
      outcome, or whether the measurement failed {(see Remark\ref{Rmk:Bell_meas})}.
    \end{itemize}
  \item Alice and Bob publicly announce the intensities they have used for all the rounds $i \in[N]$
  (the order in which this is announced is not important). Alice checks that
  among the rounds where she has used intensity $a_s$ \emph{and} the measurement succeeded, the fraction $f_{b_s}$ of
  rounds where Bob has used intensity $b_s$ is higher than $p_{b_s} - \beta^A$. Bob checks that
  among the rounds where he has used intensity $b_s$ \emph{and} the measurement succeeded, the
  fraction $f_{a_s}$ of rounds where Alice has used intensity $a_s$ is higher than $p_{a_s} - \beta^B$.
  If this is not the case, Alice or Bob abort the protocol.
  \item Using the decoy states Alice estimates a lower-bound $L_{A1}$ for $n_1^A$ (this is given by Lemma \ref{Thm:estimation}), the number of rounds where the Bell
   measurement has not been announced as a failure
  \emph{and} where Alice emitted $1$ photon with intensity $a_s$.
  If $\frac{U_{A2}}{f_{b_s}(n_1^A+n_{\geq 2}^A)}\geq \gamma + \alpha_4^A$
  Alice aborts the protocol.
  \item Using the decoy states Bob estimates a lower-bound for $n_1^B$ (this is given by Lemma \ref{Thm:estimation}), the number of rounds where the Bell
   measurement has not been announced as a failure
  \emph{and} where Bob emitted $1$ photon with intensity $b_s$.
  If $\frac{U_{B2}}{f_{a_s}(n_1^B+n_{\geq 2}^B)}\geq \gamma +\alpha_4^B$
  Bob aborts the protocol.
  \item Alice and Bob discard all the rounds where a failure has been announced, and where the intensities
  used by Alice and Bob are not $a_s$ and $b_s$. Let's call the remaining number
  of rounds $n$. Alice has strings $X_1^n$ and $\Theta_1^n$ $\in \{0,1\}^n$, and Bob has strings
  $\hat X_1^n$ and $\hat \Theta_1^n$ $\in \{0,1\}^n$. Note that $n=f_{b_s} \times (n_1^A+n_{\geq 2}^A)= f_{a_s} \times (n_1^B+n_{\geq 2}^B)$.
  Alice and Bob check that $n\geq \frac{l+2\log(1/2\epsilon)+\ln(\epsilon^{-1})}{\lambda-h(\delta)}$, and otherwise abort the protocol.

  \item Both parties wait for a time $\Delta t$.
  \item Alice sends $\Theta_1^n$ over to Bob.
  \item Bob computes the set $\mathcal{I} \subseteq [n]$ of rounds $i$ where $\Theta_i= \hat \Theta_i$.
  Bob discards all the rounds $j \notin \mathcal{I}$. Let's then
  call $\hat X_\mathcal{I}$ the string formed by all the remaining bits $\hat X_i$ with $i\in \mathcal{I}$.

  Note that when there
  is no noise we have that $\forall i \in \mathcal{I}$ $X_i=\hat X_i$. In practice there are always errors: We will
  call $e_{\rm err}$ the expected errors rate between $X_i$ and $\hat X_i$ (for $i\in \mathcal{I}$).

  \end{enumerate}
  \item[Commit Phase]\hfill\\\vspace{-1em}
  \begin{enumerate}
    \item Bob checks whether $m:=|\mathcal{I}| \in [1/2 \cdot n - \alpha_1,1/2 \cdot n + \alpha_1]$. If this is not the case Bob
    aborts.
    \item Alice chooses a random $[n,k,d]$-linear code $\mathcal{C}$ (for fixed $n$ and $k$) and computes $w=\textsf{Syn}(X_1^n)$ and sends it to Bob.
    \item Alice picks a random 2-universal hash function $r \in_R \mathcal{R}$ and sends it to Bob.
    \item Alice outputs $C_1^l:={\rm Ext}(X_1^n,r)$ where ${\rm Ext(\cdot,\cdot)}$ is a randomness extractor
    from the 2-universal family of function.
  \end{enumerate}

  \item[Open phase]\hfill\\\vspace{-1em}
  \begin{enumerate}
    \item Alice sends $X_1^n$ to Bob.
    \item Bob computes its syndrome and checks if it agrees with $w$ he received from Alice in the Commit phase. If they disagree
    Bob aborts.
    \item Bob checks that the number of rounds $i\in \mathcal{I}$ where $X_1^n$ and $\hat X_\mathcal{I}$ do not agree lies in
    the interval $]e_{\rm err}- \alpha_2, e_{\rm err} +\alpha_2[$. If not, Bob aborts the protocol, otherwise
    he outputs $C_1^l:={\rm Ext}(X_1^n,r)$ where ${\rm Ext(\cdot,\cdot)}$ is a randomness extractor
    from the 2-universal family of function.
  \end{enumerate}
\end{description}

\end{Ptol}
\end{framed}

\renewcommand*{\thefootnote}{\fnsymbol{footnote}}
We require that $(p-\sqrt{\ln(\epsilon^{-1})/2N})\, N\geq n^\star,$ for $n^\star$ {satisfying} $ {({\lambda-h(\delta)})} n \geq {l+2\log(1/2\epsilon)+\ln(\epsilon^{-1})}$\footnote[3]{Note
 that $\lambda$ and $\delta$ implicitly depend on $n$, therefore one cannot solve
the inequality analytically.} only
to make sure there are enough rounds to produce $l-$bits final strings in a secure way.
$p$ is the probability that a round is not discarded in the honest scenario, and it can be expressed
a function of the experimental parameters: The round won't be discarded if
both players sent a signal state for this round, which happens with probability
$p_{a_s}\times p_{b_s}$, and if the measurement station did not reported this
round as failure {(see Remark \ref{Rmk:Bell_meas})} which happens with probability $1-p_{\rm fail|a_s b_s}$,
so $p=p_{a_s}\times p_{b_s}\times (1-p_{\rm fail|a_s b_s})$.

\renewcommand*{\thefootnote}{\arabic{footnote}}
\subsection{OT with an imperfect single photon sources}
\label{Sec:OT_imperfect}

In this section we will prove that MDI Oblivious Transfer is ``not easy'' in practical settings.
Indeed in practice photon sources are not perfect \textit{i.e.}~they have some probability
$p_{\geq2}$ to emit more than one photon. If now one considers a protocol containing a preparation phase
similar to the one of Protocol \ref{Ptol:OT}, but where now Bob has an imperfect single
photon source, it becomes possible for a malicious Alice to
deduce from the states she receives from Bob, some of the bases $\Theta_i$ that
have been used in Bob's encoding. As we will explain below this
is due to the fact that when more than one photon are emitted by Bob's source,
a dishonest Alice can distinguish states encoded in the standard and the Hadamard basis, which
is not possible to do when a single photon is emitted. This is a leakage
of information that has heavy consequences on the feasibility of an OT protocol as explained below.\\

We will illustrate how this leakage of information can break security of a protocol,
by describing what happens to Protocol \ref{Ptol:OT} when Alice is malicious
and Bob holds an \emph{imperfect} single photon source. After this we will generalize the reasoning.

Dishonest Alice's end goal is to guess correctly the value of bit $C$ that Bob will get
at the end of the protocol. Moreover, Alice being malicious implies that Alice has
full control over the measurement station, and therefore everything
Bob sends to the measurement station can be considered in Alice's possession.
Let us now start with the preparation phase of Protocol \ref{Ptol:OT}.
In this phase of the protocol, Bob sends BB84 states\footnote{$\ket{X_i}_{\Theta_i}$ which correspond to
encoding in the basis $\hat \Theta_i$, where $\hat \Theta_i=0$ corresponds to the standard basis and $\hat \Theta_i=1$ corresponds to the
Hadamard basis} to the measurement station,
or equivalently to dishonest Alice. But contrary to section \ref{Sec:OT_perf} Bob now holds
an \emph{imperfect} single photon source. This means that in some of the rounds, more than
one photon are sent to Alice. This becomes a problem because if, for example, the source has emitted two photons,
 then the state Alice receives conditioned on Bob preparing it in the
standard basis is $1/2(\ketbra{00}{00} + \ketbra{11}{11})$, while if we condition the state
on being prepared in the Hadamard basis it is $1/2(\ketbra{++}{++} + \ketbra{--}{--})$. These
two states are not the equal, and therefore Alice can use these states to guess the basis $\Theta_i$ that Bob has used
to encode the state. When a single photon is used this is not a problem since
$1/2(\ketbra{0}{0}+\ketbra{1}{1}) = \id/2 = 1/2(\ketbra{+}{+}+\ketbra{-}{-})$: the two cases -- Bob prepares the
state in the standard or the Hadamard basis -- are perfectly indistinguishable.
Moreover, the more photons are emitted by the source, the easier it is for Alice to
guess correctly which basis Bob has used. To be conservative, for each round in which
multiple photons have been emitted we will consider that malicious Alice knows exactly Bob's choice of basis
$\hat \Theta_i$.

At the end of the preparation phase malicious Alice
sends a string $\Theta_1^n$\footnote{The way Alice chooses the value for $\Theta_1^n$ has no importance, and we will
therefore consider $\Theta_1^n$ as a fully random string in this argument.} to Bob. Bob uses the string $\Theta_1^n$
he received from Alice and his own choice of bases described by the string $\hat \Theta_1^n$
to compute the set $\mathcal{I}:=\{i\in [n] : \Theta_i = \hat \Theta_i\}$, which is the
set of rounds in which Bob's choice of bases matches the value of the bit malicious Alice has sent to him,
and where $n$ denotes the total number of rounds. He also
erases all the bits $\hat X_i$ he has used to encode the states he has sent to the station for all $i$ such that $i \not\in \mathcal{I}$.
At this point Bob holds the set $\mathcal{I}$ and the string $\hat X_{\mathcal{I}}$ which is formed
by all the bits $\hat X_i$ he has used in the round $i \in \mathcal{I}$.
Remember that Malicious Alice knows the value of $\hat \Theta_i$ in some of the rounds,
and therefore knows whether these rounds correspond to rounds in $\mathcal{I}$ or not. We call $I_G$
the set of rounds for which Alice knows that they are in $\mathcal{I}$ and $I_B$ the set of rounds
for which she knows that they are not in $\mathcal{I}$. The choice bit $C$ that has to be created by the
protocol is chosen uniformly at random by Bob. He then uses this bit $C$ to rename the sets
$(\mathcal{I},\mathcal{I}^c)$ -- where $\mathcal{I}^c$ denotes the complement of $\mathcal{I}$ --
into $(I_C,I_{1-C})$, where $C$ takes value in $\{0,1\}$. In other words, if Bob chooses
$C=0$ then $(\mathcal{I},\mathcal{I}^c)$ is renamed into $(I_0,I_{1})$, and if he chooses $C=1$,
$(\mathcal{I},\mathcal{I}^c)$ is renamed into $(I_1,I_{0})$.

After the preparation phase, Bob sends $(I_0,I_{1})$ to (malicious) Alice.
Revealing these two sets to Alice does not reveal in itself the value of bit $C$. However,
there has been a leakage of information in the preparation phase, and from this leakage Alice
knows the set $I_G$ and $I_B$ defined above, she can compare the two sets $I_G$ and $I_B$ with
the sets $I_0$ and $I_1$ she has received from Bob. But by definition of $I_G$ we must have
$I_G \subset \mathcal{I}=I_C$ and $I_G \cap \mathcal{I}^c = \emptyset$. Therefore she can get the value
of $C$: if $I_G \subset I_0$ then $C=0$, and if $I_G \subset I_1$ then $C=1$. Therefore Protocol \ref{Ptol:OT}
is not secure if Bob holds an imperfect single photon source.

In the following we will generalize the settings to show that the argument presented above holds for more
general protocols than Protocol \ref{Ptol:OT}. To do so we will abstract the structure of the protocol, as well as the meaning
of the registers (e.g.~the registers $I_G$ and $I_B$) we use in the attack. The notation will
stay very similar to what we have presented above, and the main intuition behind the attack remains the same.
Our impossibility result holds for any protocol satisfying
Assumption \ref{Hyp:imposs_informal}.\\

The statement we will make is expressed in in terms of asymptotic security,
\textit{i.e.}~we will say that Alice can cheat if she has a non-negligible advantage in guessing
Bob's bit $C$ (see Theorem \ref{Thm:Alice_cheat} below). A function is said to be negligible (in some variable $n$) if
it is smaller than $1/n^a$ (for any $a>0$ and for $n$ large enough). Similarly
we will say that a probability $p$ is overwhelmingly {large} if $1-p$ is negligible.

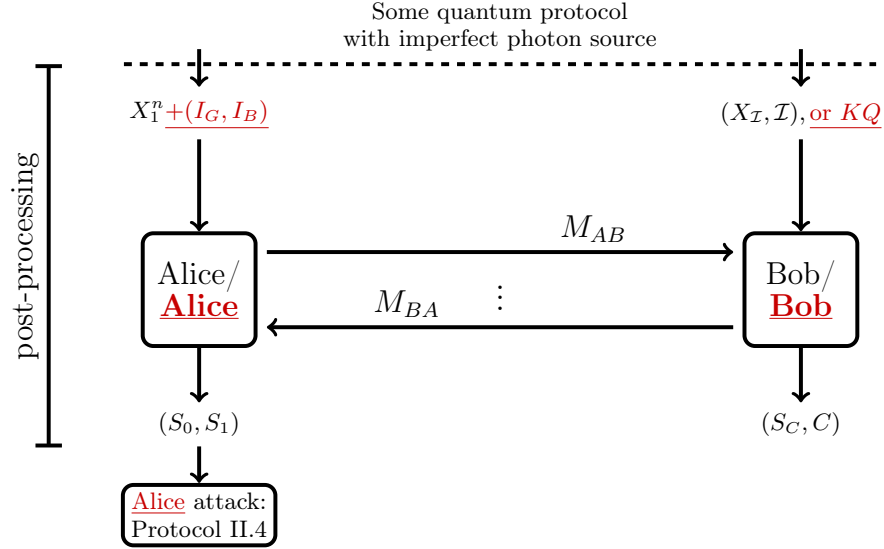
\begin{figure}[h!]
  \begin{tikzpicture}[line width=1.5pt,scale=1.0]
    \node[align=center] at (0,3.5) {Some quantum protocol\\ with imperfect photon source};
    \draw[->] (-4,3.2) -- +(0,-0.5);
    \draw[->] (4,3.2) -- +(0,-0.5);
    \draw[-,dashed] (-5,3) -- (5,3);
    \node (Alice) [draw,rectangle,rounded corners,minimum width=1.5cm,minimum height=1.5cm] at (-4,0) {$\substack{\text{\large Alice/}\\\textcolor{darkred}{\underline{\textbf{\large Alice}}}}$};
    \node (Bob) [draw,rectangle,rounded corners,minimum width=1.5cm,minimum height=1.5cm] at (4,0) {$\substack{\text{\large Bob/}\\\textcolor{darkred}{\underline{\textbf{\large Bob}}}}$};
    \draw[->] (Alice)+(0,2)node[above]{$X_1^n\textcolor{darkred}{\underline{+(I_G, I_B)}}$} -- (Alice);
    \draw[->] (Bob)+(0,2)node[above]{$(X_{\mathcal{I}}, \mathcal{I}), \textcolor{darkred}{\underline{{\rm or}\ KQ}}$} -- (Bob);
    \draw[->] (-3.1,0.5) --node[pos=0.7,above]{\large $M_{AB}$} (3.1,0.5);
    \draw[->] (3.1,-0.5) --node[pos=0.7,above]{\large $M_{BA}$} (-3.1,-0.5);
    \node at (0,-0){\Large $\vdots$};
    \draw[->] (Bob) -- (4,-1.5)node[below]{$(S_C,C)$};
    \draw[->] (Alice) -- (-4,-1.5)node[below](Out_A){$(S_0,S_1)$};
    \draw[|-|] (-6,3) -- (-6,-2.1)node[pos=0.5,above,rotate=90,align=center] {\large post-processing};
    \node[draw,rounded corners, align=center] (Alice_attack) at (-4,-3) {\textcolor{darkred}{\underline{Alice}} attack:\\ Protocol \ref{Ptol:dishonest_Alice_OT}};
    \draw[->] (Out_A) -- (Alice_attack);

  \end{tikzpicture}
  \caption{Schematic view of the classical post-processing between (\textcolor{darkred}{\underline{dishonest}}) Alice and  (\textcolor{darkred}{\underline{dishonest}}) Bob.
  Before the post-processing Alice and Bob have run an unspecified quantum protocol which gave them their inputs: $X_1^n\textcolor{darkred}{\underline{+(I_G, I_B)}}$ for
  (\textcolor{darkred}{\underline{dishonest}}) Alice, and $(X_{\mathcal{I}}, \mathcal{I}), \textcolor{darkred}{\underline{{\rm or}\ KQ}}$ for (\textcolor{darkred}{\underline{dishonest}})
  Bob.\newline
  When Alice is dishonest, we will consider that she is ``honest but curious'' at the post-processing level, meaning that Alice will run the post-processing honestly with Bob, but she
  will record all communication $M_{AB}, M_{BA}$ and use them at the end together with her extra-input $\textcolor{darkred}{\underline{(I_G, I_B)}}$, to extract more information than
  what she should get out of the protocol. To do so she will use the strategy described in Protocol \ref{Ptol:dishonest_Alice_OT}}
  \label{Fig:equivalence}
\end{figure}

In order to generalize the attack on Protocol \ref{Ptol:OT} we have seen above,
we work in a model (see Fig.~\ref{Fig:equivalence}) where Alice and Bob have already run a quantum phase of a protocol,
that has given registers $X_1^n$ to honest Alice and $X_{\mathcal{I}},\mathcal{I}$ to Bob. $X_1^n$ is
a bit string and $X_{\mathcal{I}}$ is a substring of $X_1^n$ whose bits are the ones corresponding the set of indices
$\mathcal{I}\subseteq [n]$. One can typically think of a ``quantum phase'' as being the preparation phase of
Protocols \ref{Ptol:OT} \& \ref{Ptol:BC_imperfect_source} for example.

If Alice is dishonest we assume
that she has recorded -- during this quantum phase -- information leaked by the imperfection of Bob's source.
We model this leakage of information
by giving dishonest Alice two extra registers $(I_G,I_B)$ that correspond to two sets of indices correlated with $\mathcal{I}$.
When Bob is dishonest we simply assume that
he holds the cq-registers $KQ$ such that his min-entropy on Alice's string $X_1^n$ is
smaller than honest Bob's one. Since we work in the bounded storage model
we assume $\log {\dim}Q \leq D$.

After this quantum phase of the protocol, we assume that Alice and Bob {perform} a
classical post-processing. One such post-processing is the
post-processing of Protocol \ref{Ptol:OT}. When a party is dishonest we assume he will in fact be semi-honest during
the post processing, meaning that
he will run the post-processing honestly but record all the information
he has received or sent.
We prove that if such a protocol is correct and secure against dishonest Bob, then
Protocol \ref{Ptol:dishonest_Alice_OT} gives dishonest Alice a (semi-honest) strategy to
use her extra input registers $(I_G,I_B)$ she got from the quantum phase and
all the communication she recorded during the post-processing in order
to guess honest Bob's output bit $C$ with non-negligible advantage.

We will describe the set of messages going from Bob to Alice by the random variable $M_{BA}$. The messages
from Alice to Bob will be described by the random variable $M_{AB}$. The random variable composed of these two
variables will be called $M$. In other words $M:=(M_{AB},M_{BA})$.

The output of honest Alice is $(S_0,S_1):=(f_0(X_1^n,M), f_1(X_1^n, M)) \in \{0,1\}\times\{0,1\}$, where $f_0$ and $f_1$ are two
functions determined by the protocol. Typically, these functions are the composition of
error correction with a randomness extractor. The output of honest Bob is $(C, S_C):=(g(X_{\mathcal{I}}, \mathcal{I},M), \tilde g(X_{\mathcal{I}}, \mathcal{I},M))$,
where $g$ and $\tilde g$ are two other functions determined be the protocol. These four functions model the operations
that honest Alice and Bob have to perform according to the protocol they are running.

We construct an attack where Alice is semi-honest (or equivalently ``honest but curious''), that is, she will execute the
post-processing part of the protocol honestly but keep all the information that she has exchanged with Bob so that
she can in the end compute whatever she is interested in, which in this case is $C$. Our result holds under two assumptions stated below.
This restricts the applicability of our theorem. However, we argue in the Discussion Section
that these assumptions should still be sufficiently general for {many} practical {settings}.

\begin{Hyp}[Informal] \label{Hyp:imposs_informal} In order to prove the theorem below we need two assumptions.
  {Let $f_0$ and $f_1$ be the functions that map honest Alice's available information $(X_1^n, M)$ to
  her outputs $S_0$ and $S_1$: $S_0:=f_0(X_1^n,M)\ \&\ S_1:=f_1(X_1^n,M)$. Let $(I_G,I_B)$ be the sets of indices
  that dishonest Alice gets before the execution of the post-processing procedure due to the
  imperfection of Bob's photon source (see Fig.~\ref{Fig:equivalence}).}
  \begin{enumerate}
    \item {There exists a computable function $F$ than maps $(X_1^n,M)$ to the pair
    of sets $(I_0,I_1)$\footnote{Remember that in
    Protocol \ref{Ptol:OT} sets $(I_0,I_1)$ correspond to the renaming of the
    sets $(\mathcal{I},\mathcal{I}^c)$ that Bob sends to Alice} corresponding to
    the positions of the bits of $X_1^n$ on which the functions $f_0$ and $f_1$ depend.}
    \item There is a non-negligible probability that,
    \begin{align*}
      \begin{cases}
        \text{the intersection between the set $I_G \cup I_B$ and
        $I_0 \backslash I_1$ is not empty}\\
        \text{and}\\
        \text{the intersection between the set $I_G \cup I_B$ and
        $I_1 \backslash I_0$ is not empty}.
      \end{cases}
    \end{align*}
    If we define $\kappa$ being the minimum size of the two intersections above, we can rephrase this condition by
    saying that, there is a non-negligible probability that $\kappa\geq1$.
   \end{enumerate}
   The sets $I_G, I_B$ are the sets dishonest Alice gets from the leakage of the quantum part of the
   protocol. The sets $I_0, I_1$ are the sets correlated to set $\mathcal{I}$ and bit $C$ that do not reveal value of bit $C$ as long
   as $\mathcal{I}$ is completely unknown from Alice. Of course since dishonest Alice has extra information $I_G, I_B$ correlated to
   $\mathcal{I}$, Alice is \emph{not} ignorant about $\mathcal{I}$: She therefore
   has some information about the bit $C$, which as we will see allows her to cheat. The reader can find a more formal
   version of these assumptions in the Methods Section:
   Assumption \ref{Hyp:assumptions_imposs}.
\end{Hyp}

\begin{Thm*}[Dishonest Alice cheating (Informal)]
  If a quantum protocol between Alice and Bob that implements OT is such that it leaks some information $(I_G,I_B)$ to dishonest Alice
  in the quantum phase (before the classical post-processing), and if this protocol is correct and secure against dishonest Bob, then there exists
  a strategy for dishonest Alice that allows her to cheat, \textit{i.e.}~she can guess Bob's bit $C$ with non-negligible advantage.
  This strategy runs as follows: Dishonest Alice runs honestly the post processing phase with Bob, but records
  all messages sent and received during this post-processing. At the end of the post-processing she will use all this messages together with
  her extra information $(I_G,I_B)$ in order to locally run the procedure described in Protocol \ref{Ptol:dishonest_Alice_OT}. This procedure
  outputs her guess for Bob's bit $C$.
\end{Thm*}

\noindent The reader can find a formal version of this theorem in the Methods Section, together with its proof: Theorem \ref{Thm:Alice_cheat}.
\newline

We recall that when Alice is dishonest she holds some extra set $I_G$ (and $I_B$), which in a protocol like Protocol
\ref{Ptol:OT} would typically
correspond to the multiphoton rounds where dishonest Alice has inferred that Bob has used the same basis as she did (or a different basis for $I_B$).
So at the end of the post-processing she will execute the strategy detailed in Protocol \ref{Ptol:dishonest_Alice_OT},
where she starts by computing the two sets $I_0$ and $I_1$. She will then choose
uniformly at random -- thanks to random bit $r$ -- whether she later wants to sample at random an index in $\mathcal{S}_0:=I_0\backslash I_1 \cap (I_G\cup I_B)$ or
in $\mathcal{S}_1:=I_1\backslash I_0 \cap (I_G\cup I_B)$. At this point Alice samples uniformly at random
an index in $\mathcal{S}_r$, and checks whether this round is
in $I_G$ or in $I_B$. If it is in $I_G$ then Alice's guess for Bob's bit $C$ will be $r$, and otherwise
she guesses $1-r$.
More formally Alice proceeds as follows.
\begin{framed}
\begin{Ptol}[Dishonest Alice's strategy]\label{Ptol:dishonest_Alice_OT}\hfill\\
  \indent {\rm \textbf{Inputs:}} $x_1^n,m,I_G, I_B$.\\
  \indent {\rm \textbf{Outputs:}} $b$.
  \begin{itemize}
  \item Alice computes $(I_0,I_1)=F(x_1^n,m)$, {where $F$ is given by Assumptions \ref{Hyp:imposs_informal}}.\\
  \item Alice checks that $I_0\backslash I_1 \cap (I_G\cup I_B) \neq \emptyset$ and $I_1\backslash I_0 \cap (I_G\cup I_B)\neq \emptyset$.
  If this is not the case Alice outputs $b \in_R \{0,1\}$ uniformly at random, otherwise she continues with the protocol.
  \item Alice sample a bit $r$ uniformly at random.
  \item Alice chooses an index $i_r \in I_{r}\backslash I_{1-r} \cap (I_G\cup I_B)$ uniformly at random.
  \item Alice checks whether $i_r \in I_G$ or $i_r \in I_B$. If $i_r \in I_G$ then Alice outputs $b=r$ and
  she outputs $b=1-r$ otherwise.
\end{itemize}
Alice's output bit $b$ represents Alice's guess for the bit $C$ that honest Bob got from the protocol.
\end{Ptol}
\end{framed}
Intuitively the sets $I_0, I_1$ carry information about the correlations Alice and Bob
share at the beginning of the post-processing, but not about Bob's final output $C$. In particular these sets say that if
their initial (honest) inputs are such that Bob knows the bits of $X_1^n$ on positions given by $I_0$ then $C=0$,
and if he initially knows the bits of $X_1^n$ on positions given by $I_1$ then $C=1$. However
since honest Alice does not know which bits of $X_1^n$ Bob knows (she is ignorant about $\mathcal{I}$), it does not
say anything about the actual value of Bob's output $C$. Dishonest Alice however gets
extra inputs $(I_G,I_B)$ that precisely gives her information about which are the bits Bob knows.
As a consequence by cross-referencing these two pieces of information dishonest Alice can get some advantage
in guessing bit $C$.

 \section{Discussion}

In the previous section we show that all protocols that satisfy the two assumptions given in
Assumptions \ref{Hyp:imposs_informal} (or more formally Assumption \ref{Hyp:assumptions_imposs})
cannot be secure against dishonest Alice. We believe that the class
of protocols that satisfy these conditions is general enough to encompass {many} of
the protocols that are currently implementable with current technology. In this section
we argue in this direction.

We first point out that it should not be possible to get a fully general impossibility
theorem, since we have shown that when having a sufficiently good
single photon source it is possible to devise a secure protocol (see Theorem \ref{Thm:Sec_OT_perf}).
As a consequence one can only prove statements about more restrictive classes of protocols. This is what we have done
in the previous section. However we have analyzed these protocols under Assumptions \ref{Hyp:imposs_informal},
and it is not clear how restrictive Assumptions \ref{Hyp:imposs_informal} are.

First, let us spell out some of the implicit assumptions made for our theorem that necessarily
limit the range of its applicability. In the model we use
(see Fig.~\ref{Fig:equivalence}), it is clear that the classical post-processing operated by
Alice and Bob runs on bit strings ($X_1^n, X_{\mathcal{I}},\ldots$) and of sets of indices ``$\mathcal{I},\ldots$'', however we think
that the reasoning used for our theorem can be extended to more general inputs. In this model, we also only
start by looking at the attack directly at a post-processing part of the protocols. This is convenient
since it allows our theorem to be valid for various quantum implementations that could
have run before the post-processing. Of course this assumes that the protocols
end with a fully classical post-processing phase. As a consequence our proof only applies
for such protocols. However, even though these
implicit assumptions limit the applicability of our theorem, we believe that this is enough for any practical implementation.
\newline

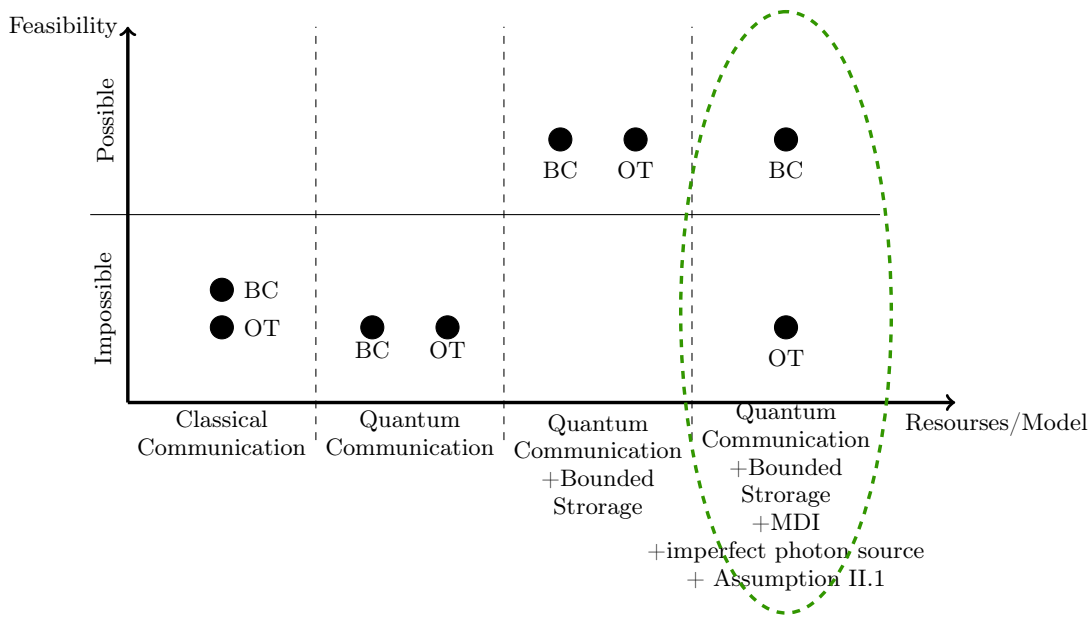
\begin{figure}[h]
  \begin{tikzpicture}[scale=1]
    \draw[->, line width=1.3pt] (0,0) -- (0,5)node[pos=1,above,left]{Feasibility};
    \draw[->,line width=1.3pt] (0,0) -- (11,0)node[pos=1.05,below]{Resourses/Model};
    \foreach \x in {1,2,3}
      \draw[-,dashed] (2.5*\x,-0.5) -- (2.5*\x,5);
    \draw[-] (-0.5,2.5) -- (10,2.5);
    \node[rotate=90] at (-0.3,1.25+2.5){Possible};
    \node[rotate=90] at (-0.3,1.25){Impossible};
    \node[align=center] at (1.25,-0.4){Classical \\ Communication};
    \node[align=center] at (1.25+2.5,-0.4){Quantum \\ Communication};
    \node[align=center] at (1.25+5,-0.85){Quantum \\ Communication\\+Bounded \\Strorage};
    \node[align=center] at (1.25+7.5,-1.25){Quantum \\ Communication\\+Bounded \\Strorage\\ +MDI\\+imperfect photon source\\+ Assumption \ref{Hyp:imposs_informal}};
    \node[draw,draw, shape=circle,fill] (dot_BC1) at (1.25,1.5) {};
    \node[right=5pt]at (dot_BC1){BC};
    \node[draw,draw, shape=circle,fill] (dot_OT1) at (1.25,1) {};
    \node[right=5pt]at (dot_OT1){OT};
    \node[draw,draw, shape=circle,fill] (dot_BC2) at (1.25+2.5-0.5,1) {};
    \node[below=2pt]at (dot_BC2){BC};
    \node[draw,draw, shape=circle,fill] (dot_OT2) at (1.25+2.5+0.5,1) {};
    \node[below=2pt]at (dot_OT2){OT};
    \node[draw,draw, shape=circle,fill] (dot_BC3) at (1.25+5-0.5,1+2.5) {};
    \node[below=5pt]at (dot_BC3){BC};
    \node[draw,draw, shape=circle,fill] (dot_OT3) at (1.25+5+0.5,1+2.5) {};
    \node[below=5pt]at (dot_OT3){OT};
    \node[draw,draw, shape=circle,fill] (dot_BC4) at (1.25+7.5,1+2.5) {};
    \node[below=5pt]at (dot_BC4){BC};
    \node[draw,draw, shape=circle,fill] (dot_OT4) at (1.25+7.5,1) {};
    \node[below=5pt]at (dot_OT4){OT};
    \draw[dashed, line width =1.3pt, color=darkgreen] (1.25+7.5,1.2) circle [x radius=4cm, y radius=14mm, rotate=90];
  \end{tikzpicture}
  \caption{Schematic representation of the ``Feasibility'' of OT and BC depending on the resources/model
  used in the protocol. In the first column neither BC nor OT are possible, but since OT can be used in order to get BC but not the contrary, OT is somewhat
  a harder problem, which is why it is below BC. When quantum communication are possible then OT and BC are equivalent (represented at the same level)
  but still impossible. In the third column we add the Bounded Storage assumption, which makes both protocol possible. They are still equivalent.
  In the last column we add that quantum communication between the parties are made in the MDI settings, and we assume that the parties
  do not have perfect single photon source. In this case BC is possible (see Theorem \ref{Thm:security_BC_imperfect}) but OT is not
  (see Theorem \ref{Thm:Alice_cheat}).}
  \label{Fig:Feasibility_Model}
\end{figure}

Let us now go to the core of our assumptions, \textit{i.e.}~let us look at conditions given in Assumptions \ref{Hyp:imposs_informal}.
The first assumption is, informally, that there exists a way for dishonest Alice to compute, from $X_1^n$ and $M$, sets of indices
$(I_0,I_1)$ that correspond to the positions in the string $X_1^n$ where the functions $f_0$ and $f_1$ are dependent on the
value of the bits located at these positions.

The second assumption can be reformulated as follows. If, for a fraction of rounds, some information is
leaked then there is a non-negligible probability that $\kappa\geq 1$ (see Assumptions \ref{Hyp:imposs_informal}).

We now argue that these two condition are not very restrictive.
\begin{itemize}
  \item[-] Indeed we conjecture that the first assumption should always hold {in
  protocols where the basis choice relates to Bob's output $C$}: In order for the protocol to
  be correct, intuitively the set of messages exchanged in the protocol represented by the random variable $M$
  should contain the
  information that ``tells'' the functions $f_0$ and $f_1$ how they should act on the bits of $X_1^n$,
  and on which of these bits they should operate.
  This suggests that Alice can also retrieve this information, \textit{i.e.}~compute $(I_0,I_1)$. We do not
  give a formal proof of this statement, that is why it is taken as an assumption. In a protocol like Protocol \ref{Ptol:OT}
  it is clear that this condition is satisfied since Bob explicitly sends the pair $(I_0,I_1)$ to Alice.
  \item[-] If the second assumption was not satisfied
  then -- at least intuitively -- Bob is able to know (with overwhelming probability) which rounds
  leak information (multiphoton emission rounds) and therefore choose the sets $I_0$ and $I_1$ (or sufficiently influence the protocol)
  such that $\kappa=0$.
  But then Bob could effectively get an almost perfect (except with negligible probability) single photon source, by preventing any
  multiphoton emission from leaving his lab. In a protocol like the ones we have presented in the previous sections,
  Bob does not know in which rounds his source has emitted multiple photons, therefore there will be in the end with very high probability multiphoton
  rounds that are kept.
\end{itemize}
For these reasons we believe that our impossibility result applies to most (if not all) currently implementable
OT protocols.

In the presence of quantum communication, it is known that OT and BC are equivalent \cite{Crepeau92,FS09},
meaning that from one of these tasks one can build a secure protocol for the other. However the construction
used, implicitly assumes a trusted device setting, and as a consequence this construction does not necessarily
prove equivalence between OT and BC in MDI settings.
Since we prove in this work that MDI BC is secure (in the bounded/noisy quantum storage model),
if our impossibility result for OT generalizes, the MDI settings (without a single photon source)
would be the first quantum setting where one can prove security for BC but not for OT with the same adversarial model
(see Fig.~\ref{Fig:Feasibility_Model}), \textit{i.e.}~it would be a quantum setting in which OT and BC are \emph{not} equivalent.

\section{Methods}

In this section we present and prove security statement for the protocols presented in the
results sections. We start by stating theorems and lemmas
that will be useful in our proofs. Then we {prove} security for BC and OT when the honest players
have perfect photon sources. We continue by giving the security proof
for BC when the honest parties only have imperfect single photon sources.
We finally prove that a class of protocols cannot be secure for OT when using imperfect single photon
sources.

\begin{Rmk}
  For simplicity, all our statements and proofs are expressed in the Bounded Storage model,
  but can easily be extended to the Noisy Storage Model as explained in {\cite{BBCW13,RPKHW18}}.
\end{Rmk}

\subsection{Useful Lemmas and Theorems}

Here, we give useful theorems that we will use as tools
for our proofs. Before stating these theorems, we need to define (smoothed) min-entropies.
{The ``smoothness'' of the smooth-min entropy is defined relatively to the purified distance
defined as follows.}

{\begin{Def}[see \cite{T15}]
  Let $\rho$ and $\sigma$ be two non-normalized quantum states. Their purified distance is given by,
  \begin{align}
    \nabla(\sigma,\rho):= \sqrt{1-F(\sigma,\rho)},
  \end{align}
  where $F$ is the fidelity defined as,
  \begin{align}
    F(\sigma,\rho):= \left(\|\sqrt{\sigma} \sqrt{\rho}\|_1+ \sqrt{(1-\tr{\rho})(1-\tr{\sigma})}\right)^2,
  \end{align}
  where $\|\cdot\|_1$ is the Schatten $1-$norm: If $A$ is a linear operator acting on a finite dimensional Hilbert space, then
  $\|A\|_1:=\tr{\sqrt{A^\dagger A}}$.
\end{Def}}

{We define the ball $\mathcal{B}(\rho,\epsilon)$ of radius $\epsilon$ centered in $\rho$, as being the set of non-normalized quantum states
whose purified distance to $\rho$ is less or equal to $\epsilon$. One can now define the smooth min-entropy.}

\begin{Def}
Let $\rho_{AB}$ be a quantum state, and let $\epsilon\geq 0$. The $\epsilon$-smoothed min-entropy on $A$ conditioned on $B$
  is defined as
  \begin{align}
    H_{\min}^\epsilon(A|B)_{\rho}:= \sup_{\hat \rho \in \mathcal{B}(\rho,\epsilon)}\left(- \inf_{\sigma_B} \inf\{\eta \in \mathbb{R}: \hat \rho \leq 2^{\eta} \id_{A} \otimes \sigma_B \} \right),
  \end{align}
  where $\sigma_B$ ranges over the density matrices, and where $\mathcal{B}(\rho,\epsilon)$
  is the ball of radius $\epsilon$ centered
  in $\rho$. The $(\epsilon=0)$-smoothed min-entropy is simply called min-entropy and is denoted $H_{\min}(A|B)_\rho$.
\end{Def}

\begin{Thm}[Leftover Hash Lemma with smooth min-entropy  \cite{RennerThesis,TomQuantum}]\label{Thm:leftoverHmin}
Let $\rho_{A_1^nE}$ be a classical-quantum state and let ${\rm Ext(\cdot,\cdot)}:\{0,1\}^n \times \mathcal{R} \mapsto \{0,1\}^l$ be an extractor
 based on a 2-universal family of hash functions $\mathcal{R}$ from $\{0,1\}^n$ to $\{0,1\}^l$,
that maps the classical $n$-bit string $A_1^n$ into $K_A$. Then
\begin{align}
\|{\rho_{K_ARE}}-{\tau_{K_A}\otimes \rho_{RE}}\|_1\leq 2^{-\frac{1}{2}\left(H_{\min}^{\epsilon}(A_1^n|E)_{{\rho}}-l\right)}+2 \epsilon,
\end{align}
where $\tau$ denotes the maximally mixed state, and $\|\cdot\|_1$ is the Schatten $1$-norm.
\end{Thm}

We will use many times a chain rule on min-entropy stating that a conditioning quantum register cannot
decrease the entropy more than by its size expressed in qubits.

\begin{Thm}[min-entropy chain rule (\cite{RennerThesis}]\label{Thm:Chain_rule}
  Let $\rho_{XKQ}$ be a classical on $XK$, and $\epsilon \geq 0$. Then we have
  \begin{align}
    H_{\min}^\epsilon(X|KQ) \geq H_{\min}^\epsilon(X|K) - \log{ {\rm dim}(Q)}.
  \end{align}
\end{Thm}

Using the ideas from \cite{Steph_1, NJCKW12} we will use random codes to prove the security of Bit Commitment.
We give here one useful property of these random codes, which can be viewed as a
tradeoff between the minimal distance $d$ of the code and its rate $R$.
\begin{Thm}[\cite{Gal62}]\label{Thm:random_code}
  For a randomly generated $[n,k,d]$ binary linear code with rate $R:=k/n$, the minimum distance $d$ satisfies,
  \begin{align}
    \Pr(d\leq \delta n) \leq 2^{(R-C_{\delta}) n},\ \text{for } 0\leq \delta \leq 1,
  \end{align}
  where $C_{\delta}:=1-h(\delta)$, $h(x):=-x \log(x)-(1-x)\log(1-x)$ is the binary entropy,
  {and where the probability is taken uniformly over all the codes with fixed parameters $k$ and $n$}.
\end{Thm}

The following min-entropy splitting lemma intuitively states
that for a classical distribution $P_{X_0X_1Z}$, if the min-entropy (conditioned on $Z$)
on $(X_0,X_1)$ is large then it must be the case
that the random variable $X_{1-C}$ has high min-entropy too, where $C$ is a binary random
variable.

\begin{Lmm}[Min-entropy splitting \cite{Wull07,DFRSS07}] \label{Lmm:min-ent_split}
  Let $X_0,X_1,Z$ be three random variables with distribution $P_{X_0X_1Z}$. Let $1>\epsilon > 0$. If
  \begin{align}
    H_{\min}^\epsilon(X_0X_1|Z) \geq K,
  \end{align}
  then there exists a binary random variable $C$ such that,
  \begin{align}
     H_{\min}^{4\epsilon}(X_{1-C}|CZ) \geq K/2-1+2\log(1-\sqrt{1-\epsilon^2}).
  \end{align}
\end{Lmm}

Very often we will use a concentration bound called the
Hoeffding inequality.
\begin{Thm}[Hoeffding inequality \cite{Hoeffding}]\label{Thm:Hoeffding}
  Let $X_1,\ldots, X_n$ be $n$ identically and independently distributed random variables.
  If $\forall i \in [n],\ a\leq X_i \leq b$, Then
  \begin{align}
    &\Pr\left(\frac{1}{n}\sum_i X_i - \mathbb{E}\left(\frac{1}{n}\sum_i X_i\right) \geq t \right) \leq \exp\left(-\frac{2t^2 n}{(b-a)^2}\right)\\
    &\text{and}\\
    &\Pr\left(\mathbb{E}\left(\frac{1}{n}\sum_i X_i\right) - \frac{1}{n}\sum_i X_i  \geq t \right) \leq \exp\left(-\frac{2t^2 n}{(b-a)^2}\right).
  \end{align}
  As a consequence, by taking $t=\sqrt{\tfrac{(b-a)^2 \ln{\epsilon^{-1}}}{2n}}$ for some $\epsilon\in]0,1[$, we get,
  \begin{align}
    &\Pr\left(\frac{1}{n}\sum_i X_i - \mathbb{E}\left(\frac{1}{n}\sum_i X_i\right) \geq \sqrt{\tfrac{(b-a)^2 \ln{\epsilon^{-1}}}{2n}} \right) \leq \epsilon \\
    &\text{and}\\
    &\Pr\left(\mathbb{E}\left(\frac{1}{n}\sum_i X_i\right) - \frac{1}{n}\sum_i X_i  \geq \sqrt{\tfrac{(b-a)^2 \ln{\epsilon^{-1}}}{2n}} \right) \leq \epsilon.
  \end{align}
\end{Thm}

\subsection{Bit Commitment (BC) with perfect single photon sources}

In this section we present the security proof for Protocol \ref{Ptol:BC_perfect_source} which implements
BC when honest parties have perfect single photon sources. In particular
we prove Theorem \ref{Thm:Sec_BC_perfect} below. The security proof is mostly the
same as in \cite{NJCKW12,Steph_1}, the only differences are that in our Protocol \ref{Ptol:BC_perfect_source}
we are guaranteed that the sources emit single photons, so we do not need to care about multiphoton emissions,
and that because we want the security to hold even in the presence of noise, we adapt the simulator argument
of \cite{Steph_1}. More over we use a more recent lower bound \cite{O_F_S} on the min-entropy.

\begin{Thm}[Security of Protocol \ref{Ptol:BC_perfect_source}] \label{Thm:Sec_BC_perfect}
  Let $\epsilon>0$ be a security parameter, $e_{\rm err} \in [0,1/2[$ is the expected error rate of the protocol \ref{Ptol:BC_perfect_source}, and let $l \in \mathbb{N}$, $l>0$ be the length of
  the string we want to commit. Let us call
  $n$ the number of quantum communication rounds in which the measurement station has clicked in Protocol \ref{Ptol:BC_perfect_source},
  and let $\alpha_2:=\sqrt{\frac{\ln \epsilon^{-1}}{2(1/2-\alpha_1)n}}$, $\alpha_1:=\sqrt{\frac{\ln \epsilon^{-1}}{2 n}}$
  which account for statistical fluctuations.
  Let $Q$ be dishonest Bob's quantum register, $K$ his classical register, and $D$ be such that $\log {\rm dim}(Q) \leq D$.
  Let $\mathcal{C}$ be a randomly generated-$[n,k,d]$ linear code with fixed $n$ and $k$ and rate $R:=k/n$. We choose
  the rate of code $\mathcal{C}$ to be $R=\ln(\epsilon)/n + 1-h(\delta)$, where $\delta:= 2 e_{\rm err}+2 \alpha_2$. Let $\lambda:= f(-D/n) -1/n$ be
  a lower-bound on the $\epsilon$-smooth min-entropy rate of honest Alice's string $X_1^n$ conditioned on (malicious) Bobs information $KQ$,
  where $f$ is defined in eq.~\eqref{eq:def_f1}.\\

  If $n$ satisfies\footnote{Since $\delta$
  and $\lambda$ implicitly depend on $n$, one cannot analytically solve the inequality.},
  \begin{align}
     ({\lambda-h(\delta)})\, n \geq {l+2\log(1/2\epsilon)+\ln(\epsilon^{-1})},
  \end{align}
  then Protocol \ref{Ptol:BC_perfect_source} implements a $(l,3 \epsilon)-$Randomized String Commitment.
\end{Thm}
\begin{proof}
  When the two parties are honest and conditioned on not aborting, one can check
  that the protocol is correct. When the two parties are honest, they can abort in two places. Either they abort
  in the first step of the Commit phase or in the third phase of the Open phase. In the first case
  Bob aborts if $|\mathcal{I}<1/2 n -\alpha_1|$. By the definition of $\alpha_1$ and the Hoeffding inequality (see Theorem \ref{Thm:Hoeffding}), this happens
  with probability at most $\epsilon$. Similarly in step 3 of the Open phase Bob
  aborts the protocol if he observes an error rate that does not lie in the interval $[e_{\rm err} -\alpha_2 , e_{\rm err} +\alpha_2]$,
  which by Hoeffding inequality happens with probability at most $2 \epsilon$. Putting this two potential abort events
  together, the honest parties have a probability at most
  $3 \epsilon$ to abort, which proves correctness.
  \newline
  Lemma \ref{Lmm:Sec_Alice} proves that Protocol \ref{Ptol:BC_perfect_source} is $3\epsilon$-hiding.
  \newline
  Lemma \ref{Lmm:sec_BC_Bob} together with Theorem \ref{Thm:random_code} show that Protocol \ref{Ptol:BC_perfect_source} is $2\epsilon-$binding.
\end{proof}

In the following with will prove Lemmas \ref{Lmm:Sec_Alice} and \ref{Lmm:sec_BC_Bob}, which state
security for honest Alice and for honest Bob respectively.

\textbf{Security for Alice:} When Bob is dishonest we will assume that he controls the measurement
station, therefore we treat the measurement station and Bob as one single party (Fig.~\ref{fig:dishonest_Bob}). Note that
this reduces to the trusted device scenario in which Bob is dishonest \cite{Steph_1,O_F_S,NJCKW12}. As a consequence several
results from Refs.~\cite{O_F_S,NJCKW12} can be reused here.

In fact, the situation in this section is even simpler in the sense that we consider that the
honest party (Alice) has access to a perfect single photon source. This, together with
the fact that we use a lower bound \cite{O_F_S} on the min-entropy that does not depend on the specifics of the state but only on
the structure of Alice's measurements, prevents Bob
from gaining any advantage by (selectively) discarding rounds. We discuss this in more details in Appendix \ref{Sec:App_selc_discard}.

Let $f(\cdot)$ be the following function.
\begin{align}\label{eq:def_f1}
  f(x):=
  \begin{cases}
    {0} &\text{ {if} }\ {x<-1}\\
    g^{-1}(x) &\text{ if }\ -1\leq  x <1/2\\
    x &\text{ if }\ {1/2\leq x  \leq 1},
  \end{cases}
\end{align}
where $g(x):= h(x)+x-1$ and $h(x):= -x \log(x)-(1-x)\log(1-x)$ is the binary entropy.

\begin{Lmm}[from \cite{O_F_S}]\label{Lmm:min-entropy_bound}
  Let $\epsilon\geq 0$. If Alice is honest, and Bob has a bounded quantum memory $Q$ (his quantum register $Q$ has dimension at most $2^D$)
  then at the end of the preparation phase, the smooth min-entropy of Bob on Alice string is
  \begin{align}
    H^\epsilon_{\min}(X_1^n|QK)_\rho \geq \lambda n,
  \end{align}
  where $\lambda= f(-D/n) -1/n -\log(2/\epsilon^2)/n$, and $K$ is Bob's classical register.
\end{Lmm}

Since in the protocol Alice sends the syndrome of her string $X_1^n$
to Bob, we need this syndrome to be sufficiently small in order to keep
the entropy relatively high so that the protocol
is secure against dishonest Bob. On the other hand, we need the distance
of the code to be sufficiently large in order to tolerate errors that might occur
between honest Alice and honest Bob. As in Ref.~\cite{NJCKW12} we use a random code: They have sufficiently
small syndrome with high distance for our purpose, and since the honest party are not using any decoding
we do not need an efficiently decodable code.

\begin{Lmm}[Security against Dishonest Bob, similar as in Ref.~\cite{NJCKW12}] \label{Lmm:Sec_Alice}
  Let $\epsilon\in]0,1[$. Let $Q$ be Bob's quantum memory such that $\log {\rm dim}(Q)\leq D$.
   Let $\mathcal{C}$ be a random $[n,k,d]$-linear code with rate $R:=k/n$. If $n$ satisfies
  \begin{align}\label{eq:cond_R}
    \begin{cases}
    \lambda-1+R >0\\\text{and,}\\
    n\geq \frac{l+2 \log(1/2\epsilon)}{\lambda-1+R}.
  \end{cases}
  \end{align}
  If Alice is honest, then the protocol is $3\epsilon$-hiding.
\end{Lmm}
\begin{proof}
  Using Lemma \ref{Lmm:min-entropy_bound} we obtain that after the Commit phase, Bob's entropy
  on Alice's string $X_1^n$ is,
  \begin{align}
    H_{\min}^\epsilon(X_1^n|QK \textsf{Syn}(X_1^n))_\rho \geq (\lambda- 1+R) n ,
  \end{align}
  where $R$ is the rate of the code $\mathcal{C}$, \textit{i.e.}, and the length of the syndrome being $n-k=(1-R)n$.
  This together with the leftover
  hash Lemma \ref{Thm:leftoverHmin} leads us to
  \begin{align}
    \rho_{C_1^l, QK \textsf{Syn}(X_1^n)} \approx_{\epsilon'} \tau_{C_1^l} \otimes \rho_{ QK \textsf{Syn}(X_1^n)},
  \end{align}
  where $\tau_{C_1^l}$ is the maximally mixed state on $C_1^l$, and
  \begin{align}
    \epsilon' = 2 \epsilon + \frac{1}{2} 2^{-\frac{1}{2} (H_{\min}^\epsilon(X_1^n|QK \textsf{Syn}(X_1^n))-l)}.
  \end{align}
  If $\lambda- 1+R > 0$, then by choosing $n$ sufficiently large we can have $\epsilon'\leq 3\epsilon$, meaning that Protocol \ref{Ptol:BC_perfect_source} is
  $3\epsilon$-hiding.
\end{proof}

\textbf{Security for Bob:}
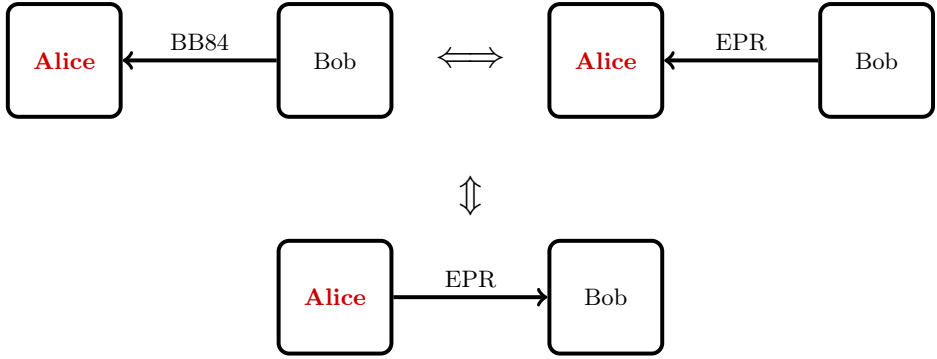
\begin{figure}[h]
  \center
  \begin{tikzpicture}[line width=1.5pt,scale=0.9]
    \node (Alice) [draw,rectangle,rounded corners,minimum width=1.5cm,minimum height=1.5cm] at (-6,0) {\textcolor{darkred}{\bf Alice}};
    \node (Bob) [draw,rectangle,rounded corners,minimum width=1.5cm,minimum height=1.5cm] at (-2,0) {Bob};
    \draw[->] (Bob) -- (Alice)node[pos=0.5,above]{BB84};
    \node at (0,0){\Large $\Longleftrightarrow$};
    \node (Alice) [draw,rectangle,rounded corners,minimum width=1.5cm,minimum height=1.5cm] at (2,0) {\textcolor{darkred}{\bf Alice}};
    \node (Bob) [draw,rectangle,rounded corners,minimum width=1.5cm,minimum height=1.5cm] at (6,0) {Bob};
    \draw[->] (Bob) -- (Alice)node[pos=0.5,above]{EPR};
    \node at (0,-2){\Large $\Updownarrow$};
    \node (Alice) [draw,rectangle,rounded corners,minimum width=1.5cm,minimum height=1.5cm] at (-2,-3.5) {\textcolor{darkred}{\bf Alice}};
    \node (Bob) [draw,rectangle,rounded corners,minimum width=1.5cm,minimum height=1.5cm] at (2,-3.5) {Bob};
    \draw[<-] (Bob) -- (Alice)node[pos=0.5,above]{EPR};
  \end{tikzpicture}
  \caption{When honest Bob has access to a single photon source and Alice is dishonest, the three situations
  depicted are equivalent: In the first Bob chooses the bases $\hat \Theta_1^n$ and $\hat X_1^n$ uniformly at random, and
  sends a BB84 type state, as described in Protocol \ref{Ptol:BC_perfect_source}. The second picture depicts
  the equivalent scenario where he sends half of an EPR pair to Alice, and gets $\hat X$ and $\hat \Theta$ by measuring the other half.
  This scenario itself is equivalent to the last {fictitious scenario} where it is dishonest Alice who sends half of the EPR pair.
  {Note that in the last fictitious scenario, sending EPR pairs might not be the optimal cheating strategy for dishonest Alice,
  but it is the one that makes this fictitious scenario equivalent to the real scenario represented in the first picture.} If some noise
  acts on the qubit sent by Bob to Alice in the first scenario, this can be seen as Alice applying a noise map on the half of the EPR
  pair she keeps before applying a measurement in the third scenario. In this virtual scenario, the other half
  of the EPR pair is assumed to be sent (measured) to (by) Bob without any noise.}
  \label{fig:dishonest_Alice_equivalence}
\end{figure}

Figure \ref{fig:dishonest_Alice_equivalence} tells us that the protocol where it is dishonest Alice that sends half of an EPR pair to
Bob produces the exact same state as Protocol \ref{Ptol:BC_perfect_source} when Alice is dishonest.
We can therefore adapt the
analysis of \cite{Steph_1} to the presence of noise
{similarly to the analysis performed in \cite{WSEE}}, which leads us to the following lemma.

\begin{Lmm}[Similar to Theorem III.5 of \cite{Steph_1} {and \cite{WSEE}}] \label{Thm:exist_X}
    If Bob is honest, then at the end of the preparation phase, there exists an ideal state $\sigma_{A \bar X_1^n \mathcal{I}}$
    between (dishonest) Alice and Bob such that:
    \begin{itemize}
      \item $\sigma_{A \bar X_1^n \mathcal{I}} = \sigma_{A \bar X_1^n} \otimes \tau_{\mathcal{I}}$
      \item $\rho_{A B} = \sigma_{A  (\bar X_{\mathcal{I}}\mathcal{I})}$,
    \end{itemize}
where $\tau_{\mathcal{I}}$ is the maximally mixed state on $\mathcal{I}$, $\rho_{AB}$ is the real state produced by the protocol between (dishonest) Alice and Bob, and where the registers
$(A,B)$ are identified with $(A, \bar X_\mathcal{I} \mathcal{I} )$.
\end{Lmm}
\begin{proof}[Proof (Sketch).]
  We will place ourselves in the virtual scenario of Figure \ref{fig:dishonest_Alice_equivalence} where Alice sends the states to Bob.
  Here, contrary to \cite{Steph_1} we want to take care of the noise that might affect the quantum signal and measurements,
  therefore the simulator introduced in Ref.~\cite{Steph_1} has to be slightly modified.

  In order to prove the existence of an ideal state $\sigma$, in Ref.~\cite{Steph_1} the authors introduce a virtual protocol where
  a simulator lies between dishonest Alice and honest Bob. This simulator
  will measure the states sent from Alice to Bob, thus creating the register $\bar X_1^n$ and then send an ``honest'' state
  to Bob. Then they show that the ideal state
  $\sigma$ created by this virtual protocol satisfies the two relations of Lemma \ref{Thm:exist_X}
  with the real state $\rho$ of the real protocol.

  In our case Fig.~\ref{fig:dishonest_Alice_equivalence} tells us that the noise will only be on the
  half of the EPR pair kept by Alice, and that the qubit sent to Bob is not affected by any noise. Therefore if the simulator
  measures it and re-encodes it honestly (and without noise) a qubit corresponding to its outcome and choice of measurement basis,
  the two relations of this lemma will be satisfied.
\end{proof}

From here on, reusing the argument in Refs.~\cite{Steph_1,NJCKW12} we get the final statement for Bob's security.
\begin{Lmm}\label{Lmm:sec_BC_Bob}
    Let $\epsilon>0$. Let $\mathcal{C}$ be an $[n,k,d]$-code with minimum distance $d$ that satisfies,
    \begin{align}
      d&\geq 2(e_{\rm err} + 2\alpha_2 )n
       \underset{n \rightarrow \infty}{\sim} 2 e_{\rm err} n,
    \end{align}
    with $\alpha_2:=\sqrt{\frac{\ln \epsilon^{-1}}{2(1/2-\alpha_1)n}}$, $\alpha_1:=\sqrt{\frac{\ln \epsilon^{-1}}{2 n}}$,
    then Protocol \ref{Ptol:BC_perfect_source} either aborts before the open phase
    or is $\epsilon$-binding according to definition \ref{Def:Sec_ideal_BC}. Note that
    the protocol specifies what the honest parties have to do when aborting. What they do during
    an abort event
    enforces security definition \ref{Def:Sec_ideal_BC} to be \emph{also} satisfied when the protocol aborts.
\end{Lmm}
\begin{proof}
  We again follow the reasoning from \cite{Steph_1,NJCKW12}. According to Lemma \ref{Thm:exist_X}
  there exists a random variable $\bar X_1^n$, such that Bob knows $\bar X_{\mathcal{I}}$ and $\mathcal{I}$.
  Now if Alice wants to cheat she needs to send to Bob a string $X_1^n \neq \bar X_1^n$ such that
  $\textsf{Syn}(X_1^n)=w$ which implies that $d_H(\bar X_1^n, X_1^n) \geq d/2$ (see \cite[Lemma IV.4]{Steph_1}), where
  $d_H(\cdot)$ is the hamming distance. Therefore Alice has to flip
  at least $d/2$ bits from $\bar X_1^n$ in such a way that $d_H(\bar X_\mathcal{I}, X_\mathcal{I}) \leq (e_{\rm err} + \alpha_2)m$.
  However Alice is ignorant about which bits Bob knows. As a consequence the situation is
  equivalent to where $\mathcal{I}$ is chosen after that Alice has chosen which bits she wanted to flip.
  This is a sampling problem, which means that we can use Hoeffding's inequality (see Theorem \ref{Thm:Hoeffding}) to estimate the number $W$
  of bits in $\mathcal{I}$ that Alice will flip:
  \begin{align}
    \Pr\big(W \leq m (d/2n - \alpha_2)\big)\leq \exp(-m \alpha_2^2) \leq \exp\left(- m \sqrt{\frac{\ln \epsilon^{-1}}{2 m}}^2 \right) \leq \epsilon. \label{eq:Hoeffding_BC}
  \end{align}
  Therefore if,
  \begin{align}
    & d\geq 2(e_{\rm err} + 2\alpha_2 )n\\
    \Rightarrow\  & m (d/2n -\alpha_2) \geq (e_{\rm err} + \alpha_2)m  \label{eq:cond_d}
  \end{align}
  then by using eqs.~\eqref{eq:Hoeffding_BC} and \eqref{eq:cond_d} we get $\Pr(W<(e_{\rm err} + \alpha_2)m)\leq \epsilon$ meaning that
  Alice's attempt in cheating is detected (and Bob will not accept) with probability $\geq 1-\epsilon$.
\end{proof}

\subsection{Oblivious Transfer (OT) with perfect single photon sources}

In this section we present and prove Theorem \ref{Thm:Sec_OT_perf} stating security for
Protocol \ref{Ptol:OT} which implements a Randomized Oblivious String Transfer when the
honest parties have access to single photon sources. The security proof
closely follows the security proofs from \cite{ENGLWW15,Steph_1}. Indeed
the main difference in our case is simply to show that security of our protocol
can be reduced to the security of \cite{ENGLWW15}. This is the case because
when Bob is dishonest, he controls the measurement station
so we are in a situation where Alice sends BB84 states to dishonest Bob, {which is
exactly the same situation as in \cite{ENGLWW15}}, and
therefore the security {immediately follows} from \cite{ENGLWW15} when Bob is dishonest. When Alice is dishonest we use
the fact that sources emit single photons together with a purification argument
in order to reduce the security of our protocol to the one of \cite{ENGLWW15}.\newline

\begin{Thm}\label{Thm:Sec_OT_perf}
  Let $\epsilon>0$ and let $l=|S_0|=|S_1|$, and $\alpha_1:=\sqrt{\frac{\ln \epsilon^{-1}}{2 n}}$.
  If the number of $n$ of quantum communication rounds in which the measurement station has clicked
  satisfies condition \eqref{eq:condition_l},
  {then} the Protocol \ref{Ptol:OT}
  implements an 1-out-2 Randomized $(l,8\epsilon)-$Oblivious String Transfer (see Def.~\ref{Def:Sec_ideal_OT}).
\end{Thm}
\begin{proof}
  Let's first check correctness with honest Alice and honest Bob. Note that conditioned on not aborting
  the protocol is $\epsilon-$correct. Indeed the only case where the protocol is not correct
  conditioned on not aborting is when the error correction procedure fails to correct
  Bob's string which happens with probability at most $\epsilon$. We then prove that when both parties are honest, the protocol aborts with probability at most $2\epsilon$.
  Indeed an abort event happens either if $|\mathcal{I}|< m$ which happens with probability at most $\epsilon$, or if
  the error correction procedure aborts which happens with probability at most $\epsilon$. As a consequence the protocol aborts
  with probability at most $2\epsilon$, and since conditioned on not aborting it is $\epsilon-$correct,
  it implies that overall the protocol is $3\epsilon-$correct.
  \newline
  According to Lemma \ref{Lmm:Sec_Alice_OT}, the protocol is $8 \epsilon-$secure for honest Alice.
  \newline
  According to Lemma \ref{Lmm:Sec_Bob_OT}, the protocol is $(\epsilon=0)-$secure for honest Bob.
\end{proof}

In the following we state and prove Lemmas \ref{Lmm:Sec_Alice_OT} and \ref{Lmm:Sec_Bob_OT} which state
security for honest Alice and for Honest Bob respectively.

\textbf{Security for Alice:} Since the preparation phase of Protocols \ref{Ptol:BC_perfect_source} and \ref{Ptol:OT}
are the same, we will use similar bounds as in Lemma \ref{Lmm:min-entropy_bound} \cite{Steph_1} to lower bound the entropy on
$X_1^n$. However we will not use the exact same bounds because we afterwards want to use the min-entropy
splitting lemma that is valid only on purely classical states. As a consequence we will
first use a chain rule (Theorem \ref{Thm:Chain_rule}) to get rid of Bob's quantum memory and then lower bound the entropy.

\begin{Lmm}\label{Lmm:Sec_Alice_OT}
  Let Bob be dishonest with a bounded quantum memory denoted $Q$ such that $\log {\rm dim}(Q)\leq D$ for some $D$. Let
  $l:=|S_0|=|S_1|$ be the length of the two strings $S_0$ and $S_1$. If
  \begin{align}\label{eq:condition_l}
    n\geq 2\frac{l+D+1-2 \log(1-\sqrt{1-\epsilon^2})}{\lambda -{\rm leak}_O -2 \alpha_1}
  \end{align}
  where ${\rm leak}_O:=|O|$ is the size of the error correction information Alice sends to Bob,
  then Protocol \ref{Ptol:OT} is $8\epsilon$-secure for Alice, with
   $\lambda= 1/2 -2 \delta'$, $\delta'=\big(2-\log(\sqrt{(32 \ln \epsilon^{-1})/n})\big) \sqrt{(32 \ln \epsilon^{-1})/n} $ \cite[eq.~(19)]{Steph_1}.
\end{Lmm}
\begin{proof}
Protocol \ref{Ptol:OT} is designed in such a way that it is sufficient to prove
that there exists a binary random variable $C$ such that the entropy $H_{\min}^\epsilon(X_{I_{1-C}}|KQCO)$
at the end of the preparation phase
is sufficiently high. Indeed after the preparation phase Alice and Bob will use a randomness extractor
on $X_{I_0}$ and on $X_{I_1}$, meaning that if the above mentioned entropy is high enough then
Bob will be ignorant of at least one of the two ``extracted'' strings, which is what we want
from the security definition.
In order to bound this entropy, we will start by bounding $H_{\min}^\epsilon(X_{I_0} X_{I_1}|KO)$ where the quantum
register $Q$ is not used, and we will reintroduce it later using a min-entropy chain rule (Theorem \ref{Thm:Chain_rule}).

Note that $X_1^n=X_{\mathcal{I}}X_{\mathcal{I}_{\rm Bad}} X_{\rm remaining}=X_{I_0} X_{I_1} X_{\rm remaining}$. By definition of $\mathcal{I}$ and
$\mathcal{I}_{\rm Bad},$ we have that $|X_{\rm remaining}|=n-2m=2\alpha_1 n$. Therefore
\begin{align}
  H_{\min}^\epsilon(X_{I_0} X_{I_1}|KO)=H_{\min}^\epsilon(X_{\mathcal{I}} X_{\mathcal{I}_{\rm Bad}}|KO) \geq H_{\min}^\epsilon(X_1^n|KO) -2\alpha_1 n.
\end{align}
By using the previous bound together with the min-entropy splitting lemma (Lemma \ref{Lmm:min-ent_split}), we get that there exists a binary random variable $C$ such that,
\begin{align}
  H_{\min}^{4 \epsilon}(X_{I_{1-C}}|KOC) \geq (H_{\min}^\epsilon(X_1^n|KO) -2 \alpha_1n) /2 -1 +2 \log(1-\sqrt{1-\epsilon^2}). \label{eq:entrp_bound_OT}
\end{align}
Using the min-entropy chain rule (Theorem \ref{Thm:Chain_rule}) on the register $Q$ ($|Q|\leq D$) and combining it with eq.~\eqref{eq:entrp_bound_OT} we conclude that
\begin{align}
  H_{\min}^{4 \epsilon}(X_{I_{1-C}}|KCO Q) &\geq H_{\min}^{4 \epsilon}(X_{I_{1-C}}|KOC)- |Q| \\
  &\geq (H_{\min}^\epsilon(X_1^n|KO) -2 \alpha_1 n) /2 -1 +2\log(1-\sqrt{1-\epsilon^2})-D,
\end{align}
where $C$ is defined by the use of the min-entropy splitting lemma in eq.~\eqref{eq:entrp_bound_OT}.

We will now again use the chain rule (Theorem \ref{Thm:Chain_rule}) to get rid of the register $O$, and we will call ${\rm leak}_O:=|O|$
the maximum leakage due to error correction, and we get
\begin{align}
  H_{\min}^\epsilon(X_1^n|KO) \geq H_{\min}^\epsilon(X_1^n|K) -{\rm leak}_O.
\end{align}
inserting this into the previous inequality gives,
\begin{align}
  H_{\min}^{4 \epsilon}(X_{I_{1-C}}|KCO Q) \geq (H_{\min}^\epsilon(X_1^n|K) -{\rm leak}_O -2 \alpha_1 n) /2 -1 +2\log(1-\sqrt{1-\epsilon^2})-D.\label{eq:fin_OT_bound}
\end{align}
The amount of error correction information ${\rm leak}_O$ sent during the protocol can be
predetermined by considering the necessary amount of error correction information the parties need when
they are \emph{both} honest, \textit{i.e.}~when both parties (and the measurement station) act in an identically and independently
distributed (IID) and trusted manner, and where
all the errors come from an i.d.d.~noise -- an ``honest noise''. Indeed if the parties are honest -- and if ${\rm leak}_O$ is
sufficiently large -- they will be able to correct their string
with probability ($\geq 1-\epsilon$), making the protocol correct. If Bob is not honest, since the amount of error correction information is
fixed, then the leakage of information is also fixed no matter what strategy he uses.
The question is now, how large is ``sufficiently large'' to allow \emph{honest} Alice and Bob to correct their
string with high probability? This question has been answered in Refs.\cite{RW05,TSSR11} where it is shown that one can take
\[
{\rm leak}_O=H_{\max}^\epsilon(X_{I_0}| \hat X_{I_0} C=0)_{\rho_{\rm honest}} +H_{\max}^\epsilon(X_{I_1}| \hat X_{I_1} C=1)_{\rho_{\rm honest}}
= 2 H_{\max}^\epsilon(X_{I_0}| \hat X_{I_0} C=0)_{\rho_{\rm honest}},
\]
where the entropies are evaluated on the state $\rho_{\rm honest}$ produced by the protocol when \emph{both} parties are honest.

One can then lower-bound $H_{\min}^{\epsilon}(X_1^n|K)$ using \cite[eq.~(19)]{Steph_1} (see also \cite{NBW12}),
\[
  H_{\min}^{\epsilon}(X_1^n|K) \geq \lambda n,
\]
with $\lambda= 1/2 -2 \delta$, $\delta=\big(2-\log(\sqrt{(32 \ln \epsilon^{-1})/n})\big) \sqrt{(32 \ln \epsilon^{-1})/n}$. Since
$H_{\max}^\epsilon(X_{I_0}| \hat X_{I_0} C=0)_{\rho_{\rm honest}}$ is evaluated on honest \emph{i.i.d}~parties we can upper-bound the max-entropy
using the equipartition Theorem \cite{TCR09}, getting $2 H_{\max}^\epsilon(X_{I_0}| \hat X_{I_0} C=0)_{\rho_{\rm honest}} \leq 2 h(e_{\rm err}) n/2 +\mathcal{O}(\sqrt{n})= h(e_{\rm err}) n +\mathcal{O}(\sqrt{n})$
where $e_{\rm err}$ is the error rate between Alice's string $X_{I_0}$ and Bob's string $\hat X_{I_0}$.

Using \eqref{eq:fin_OT_bound} and the fact that $S_{1-C}:={\rm Ext}(X_{I_{1-C}},r_{1-C})$ we can invoke the leftover hash lemma \ref{Thm:leftoverHmin} to get that,
\begin{align}
  \rho_{S_C Q S_{1-C}} \approx_{8\epsilon} \sigma_{S_C Q C} \otimes \tau_{S_{1-C}},
\end{align}
where $\tau_{S_{1-C}}$ denotes the maximally mixed state on ${S_{1-C}}$.
\end{proof}

\textbf{Security for Bob:} Once again the preparation phase is the same as for Protocol \ref{Ptol:BC_perfect_source}, therefore
we also use Lemma \ref{Thm:exist_X} to show that the protocol is secure for Bob. Intuitively this
is true because at the end of the preparation phase, Alice is ignorant about $\mathcal{I}$, and
 after that no information about $C$ is leaked.
\begin{Lmm}\label{Lmm:Sec_Bob_OT}
  If Bob is honest, then Protocol \ref{Ptol:OT} satisfies the security definition for Bob.
\end{Lmm}
\begin{proof}[Proof (Informal).]
 Since the ideal state
satisfies $\sigma_{A \bar X_1^n \mathcal{I}}= \sigma_{A \bar X_1^n}\otimes \tau_{\mathcal{I}}$ and that the only information
sent from Bob to Alice is $(I_0,I_1)$, there is no leakage on the value of $C$, therefore Alice remains ignorant about $C$. In other words
the state $\sigma_{A' S_0 S_{1} C}$ created by applying Protocol \ref{Ptol:OT} (with dishonest Alice) on the ideal state $\sigma_{A \bar X_1^n \mathcal{I}}$,
satisfies the condition $\sigma_{A' S_0 S_{1} C}=\sigma_{A' S_0 S_{1}} \otimes \tau_C$.

Also since from Lemma \ref{Thm:exist_X} $\sigma_{A (X_{\mathcal{I}} \mathcal{I})} = \rho_{A (X_{\mathcal{I} } \mathcal{I})}$ and that the
same operations are applied in the ideal and real scenario (on the registers $A X_{\mathcal{I}} \mathcal{I}$) we get that
$\sigma_{A' S_C C} = \rho_{A' S_C C}$.
\end{proof}

\subsection{Bit Commitment with an imperfect single photon sources}

In this section we present security proof for Protocol \ref{Ptol:BC_imperfect_source} which implements
String Commitment when honest parties have \emph{imperfect} single photon sources. In particular
we prove Theorem \ref{Thm:security_BC_imperfect} below. The proof is essentially the same as for Theorem
\ref{Thm:Sec_BC_perfect}, but of course since we are now dealing with imperfect single photon
sources, we need to be more careful the rounds where the honest party sends multiple photons. Indeed
in this case the malicious party could try to selectively discard the rounds where he receives less information,
typically the single photon rounds, and only keep the rounds that might leak some information, the multiphoton
rounds. To prevent that we add decoy states in the preparation phase, which will allow the honest party to check
how many multiphoton rounds are kept at the end of the preparation phase as compare to the
single photon rounds. If to many multiphoton rounds are kept at the end of the preparation
phase, the honest party aborts the protocol.\newline

\begin{Thm}\label{Thm:security_BC_imperfect}
  Let $\epsilon,\varepsilon,\hat \varepsilon, \epsilon_1$ be as defined in Lemma \ref{Thm:estimation}, and let $\gamma,e_{\rm err} \in [0,1/2[$.
  The values of the parameters $\gamma$ and $e_{\rm err}$ can be chosen by estimating the parameters honest devices.
  Let $N$ be the total number of quantum communication rounds of the preparation phase of Protocol \ref{Ptol:BC_imperfect_source}, let
  $n_k^H$ be the number of these rounds in which party $H$'s source ($H\in\{\text{Alice},\text{Bob}\}$) has produced $k$ photons \emph{and}
  in which the measurement station has clicked, and let
  $n$ be the number of communication rounds that are \emph{not} discarded at the end of the preparation phase.
  Let $\alpha_2:=\sqrt{\frac{\ln \epsilon^{-1}}{2(1/2-\alpha_1)n}}$, $\alpha_1:=\sqrt{\frac{\ln \epsilon^{-1}}{2 n}}$,
  $\alpha_1'':= \sqrt{\frac{\ln \epsilon^{-1}}{2(1-\gamma-\alpha_4^B)n}}$,
  $\alpha_1':=\min\left[1/2\,;\, \frac{\alpha_1+(1-\gamma-\alpha_4^B)\alpha_1''}{\gamma+\alpha_4^B}\right]$,
  $\alpha_3:=\sqrt{\frac{\ln \epsilon^{-1}}{2 n \big[1/2-\alpha_1'' - (1/2+\alpha_1')(\gamma+\alpha_4^B)\big]}}$.
  Let
  \begin{align*}
    \beta^A &:=\sqrt{\ln(1/\epsilon)/(2 (n_1^A+n_{\geq 2}^A))}, \text{ we will assume that $\beta^A\leq p_{b_s}/2$},\\
    \beta^B &:=\sqrt{\ln(1/\epsilon)/(2 (n_1^B+n_{\geq 2}^B))}, \text{ we will assume that $\beta^B\leq p_{a_s}/2$},\\
    \alpha_4^A &:= (2/p_{bs}+1/f_{b_s}) \beta^A,\\
    \alpha_4^B &:= (2/p_{a_s}+1/f_{a_s}) \beta^B.
  \end{align*}
  Let $\mathcal{C}$ be a randomly generated $[n,k,d]$ (with fixed $n$ and $k$) linear code with rate $R:=k/n$. We choose this
  code such that the rate $R=\ln{(\epsilon)}/n +1 - h(\delta)$, where
  $\delta:= 2\left[(1/2+\alpha_1')(\gamma+\alpha_4^B)+\alpha_3+\frac{(e_{\rm err}+\alpha_2)(1/2+\alpha_1)}{(1/2-\alpha_1')(1-\gamma-\alpha_4^B)}\right]$.
  Let $Q$ be Bob's quantum register, and let $D$ be such that $\log {\rm dim}(Q)\leq D$.
  Let $\lambda:= f(-D/n) -(\gamma+\alpha_4^A)-1/n$ lower-bound the
  $\epsilon$-smooth min-entropy rate $(H^\epsilon_{\min}(X_1^n|QK)_\rho) /n$
  except with probability $16(\epsilon+\varepsilon+\hat \varepsilon)+8\epsilon_1$ , where $f$ is defined in eq.~\eqref{eq:def_f2}.\\
  If the single photon sources used by honest parties are sufficiently good, \textit{i.e.}
  \begin{align}\label{eq:cond_thm}
    \begin{cases}
      p_{\geq 2|a_s}/(1-p_{0|a_s}) \leq p_{b_s} \gamma,\\
      {\rm{and}}\\
      p_{\geq 2|b_s}/(1-p_{0|b_s}) \leq p_{a_s} \gamma,
    \end{cases}
  \end{align}
  and if $(p-\sqrt{\ln(\epsilon^{-1})/2N})\, N\geq n^*$, where $n^*$ the smallest
   positive integer solution to the following inequality\footnote{Remember
   that the parameters like $\lambda$, $\delta$ etc. depend on $n$ },
  \begin{align}\label{eq:cond_thm}
      n\geq \frac{l+2\log(1/2\epsilon)+\ln(\epsilon^{-1})}{\lambda-h(\delta)},
  \end{align}
  and where $p$ is the probability that a round $i\in [N]$ is not discarded in the preparation phase when both parties are honest,
  then Protocol \ref{Ptol:BC_imperfect_source} implements
  a $\big(l,9 \epsilon +32(\epsilon+\varepsilon+\hat \varepsilon)+16\epsilon_1\big)-$1-out-2-Randomized String Commitment.
\end{Thm}
\begin{proof}
  Let's start with correctness. First of all note that conditioned on not aborting the protocol is correct. We now
  show that when both parties are honest the protocols aborts with probability smaller than
  $9\epsilon + 32(\epsilon+\varepsilon+\hat \varepsilon)+16 \epsilon_1$, which implies that the protocol is
  $(9\epsilon + 32(\epsilon+\varepsilon+\hat \varepsilon)+16 \epsilon_1)-$correct.
  Using the Hoeffding inequality (see Theorem \ref{Thm:Hoeffding}) it is easy to check that
  the honest parties will abort with probability at most $2\epsilon$ at step $2$ of the preparation phase.

  If the two parties are honest with sources such that
  $p_{\geq 2|a_s}/(1-p_{0|a_s}) \leq p_{b_s} \gamma$ (for Alice) and
  $p_{\geq 2|b_s}/(1-p_{0|b_s}) \leq p_{a_s} \gamma$ (for Bob), then the probability to abort at step $3$ is
  at most $\epsilon+16(\epsilon+\varepsilon+\hat \varepsilon)+8 \epsilon_1$  and at most $\epsilon+16(\epsilon+\varepsilon+\hat \varepsilon)+8 \epsilon_1$ at step
  $4$. Indeed in step 3, using the Hoeffding inequality one can check that with probability at most $\epsilon$, we have
  $\frac{n_{\geq2}^A}{n_1^A+n_{\geq 2}^A} \leq p_{b_s} \gamma + \beta^A$.
  By dividing the expression by $f_{b_s}$ and using that conditioned on not aborting in
  the previous steps $f_{b_s}\geq p_{b_s}-\beta^A$ we get
  $\frac{n_{\geq2}^A}{f_{b_s}(n_1^A+n_{\geq 2}^A)} \leq \frac{p_{b_s}}{p_{b_s} -\beta^a} +1/f_{b_s} \beta^A$.
  Using that except with probability $16(\epsilon+\varepsilon+\hat \varepsilon)+8\epsilon_1$ we have $U_{A2}\geq n_{\geq 2}^A$ and that
  for $1/p_{b_s} \beta^A \leq 1/2$ we have $1/(1-1/p_{b_s}\beta^A)\leq 1+2/p_{b_s} \beta^A$ we get the desired result. An analog proof holds
  for step 4. Again by the Hoeffding inequality, there is a probability at most $\epsilon$ to abort at step 5.

  Using again the Hoeffding inequality one can check that Bob will abort the protocol with probability
  at most $2\epsilon$ at step 1 of the Commit phase and with probability at most
  $2\epsilon$ at phase 3 of the open phase. Over all the protocol aborts with probability at most
  $9\epsilon + 32(\epsilon+\varepsilon+\hat \varepsilon)+16 \epsilon_1$.
\newline
  Security for honest Alice is given in Lemma \ref{Lmm:Sec_Alice_np}.
  Security for honest Bob is given in Lemma \ref{Lmm:Sec_Bob_np}.
\end{proof}

Before proving security for honest Alice (Lemma \ref{Lmm:Sec_Alice_np}) and for honest
Bob (Lemma \ref{Lmm:Sec_Bob_np}), we need to prove that the honest party $H \in \{Alice,Bob\}$ can always find
a lower-bound $L_{H1}$ on $n_1^H$, the number rounds where $H$ has emitted a single photon and has sent
a ``signal'' state.
This is what the following lemma shows. You can find its proof in Appendix \ref{Sec:Thm:estimation}.

\begin{Lmm}\label{Thm:estimation}
   Let $x_{o,\theta}^i$ be the ``observed'' number of
  rounds where $H$ has prepared a signal of intensity $i$ in the basis $\theta$ and where the measurement station
  (or the dishonest party) reported outcome $o\neq$failure. Let $\epsilon,\epsilon_1 >0$, and $\varepsilon, \hat \varepsilon$
  such that $\forall (o,\theta)$, $(2\varepsilon^{-1})^{1/\zeta_{o,\theta, L}} \leq \exp(3/(4\sqrt{2}))^2$ and $(\hat \varepsilon^{-1})^{1/\zeta_{o,\theta, L}}<\exp(1/3)$,
  with  $\zeta_{o,\theta, L}:=x_{o,\theta}^i-\sqrt{\sum_i x_{o,\theta}^i /2\ \ln(1/\epsilon)}$.
  Let $\Delta_{i,o,\theta}:=g(x_{o,\theta}^i, \varepsilon^4/16)$, $\hat \Delta_{i,o,\theta} :=g(x_{o,\theta}^i, \hat \varepsilon^{3/2})$, and $g(x,y):= \sqrt{2x \ln(y^{-1})}$.
  Then if $q=2$ ($q$ is the number of decoy states
  used during the protocol \textit{i.e.}~$i\in \{i_s,i_{d_1},i_{d_2}\}$)
  we have,
  \begin{align}
    n_1^H \geq L_{H1}:= \sum_{o, \theta} \left[p_{i_s|k=1}\ |S_{1,o,\theta}|_{\min} - g(p_{i_s|k=1}\ |S_{1,o,\theta}|_{\min},\epsilon_1)\right]
  \end{align}

  except with probability $16 (\epsilon+\varepsilon+\hat \varepsilon) +8\epsilon_1$, where $|S_{1,o,\theta}|_{\min}$ is given by,
  \begin{align}
    |S_{1,o,\theta}|_{\min} := \min(V_1,V_2,V_3,V_4),
  \end{align}
  with
  \begin{align}
    V_1=\frac{p_{i_{d_1}|k\geq 2}(x_{o,\theta}^{i_{d_2}} +\Delta_{i_{d_2},o,\theta})- p_{i_{d_2}|k\geq 2} (x_{o,\theta}^{i_{d_1}} + \Delta_{i_{d_1},o,\theta})}{p_{i_{d_1}|k=1} p_{i_{d_2}|k\geq 2} - p_{i_{d_1}|k\geq 2} p_{i_{d_2}|k=1} }\\
    V_2=\frac{p_{i_{d_1}|k\geq 2}(x_{o,\theta}^{i_{d_2}} - \hat \Delta_{i_{d_2},o,\theta})- p_{i_{d_2}|k\geq 2} (x_{o,\theta}^{i_{d_1}} + \Delta_{i_{d_1},o,\theta})}{p_{i_{d_1}|k=1} p_{i_{d_2}|k\geq 2} - p_{i_{d_1}|k\geq 2} p_{i_{d_2}|k=1} }\\
    V_3=\frac{p_{i_{d_1}|k\geq 2}(x_{o,\theta}^{i_{d_2}} +\Delta_{i_{d_2},o,\theta})- p_{i_{d_2}|k\geq 2} (x_{o,\theta}^{i_{d_1}} - \hat \Delta_{i_{d_1},o,\theta})}{p_{i_{d_1}|k=1} p_{i_{d_2}|k\geq 2} - p_{i_{d_1}|k\geq 2} p_{i_{d_2}|k=1} }\\
    V_4=\frac{p_{i_{d_1}|k\geq 2}(x_{o,\theta}^{i_{d_2}} -\hat\Delta_{i_{d_2},o,\theta})- p_{i_{d_2}|k\geq 2} (x_{o,\theta}^{i_{d_1}} -\hat\Delta_{i_{d_1},o,\theta})}{p_{i_{d_1}|k=1} p_{i_{d_2}|k\geq 2} - p_{i_{d_1}|k\geq 2} p_{i_{d_2}|k=1} } .
  \end{align}

  One can compute tighter bounds using more decoy states (\textit{i.e.}~for $q>2$). For more details see Appendix \ref{Sec:Thm:estimation}.
\end{Lmm}

For simplicity we will, in the following, continue the security analysis for the case where $q=2$.
The above lemma will allow us to prove the following security lemmas: Lemma \ref{Lmm:Sec_Alice_np}
proves security for honest Alice, and Lemma \ref{Lmm:Sec_Bob_np} proves security for honest Bob.

\textbf{Security for Alice:} When Alice is honest almost nothing changes except that
Bob's entropy about Alice's string is smaller by roughly $\gamma n$ bits. As a consequence
lemma \ref{Lmm:min-entropy_bound} has to be changed.

Let $f(\cdot)$ be the following function.
\begin{align}\label{eq:def_f2}
  f(x):=
  \begin{cases}
    {0} &\text{ {if} }\ {x<-1}\\
    g^{-1}(x) &\text{ if }\ -1 \leq  x <1/2\\
    x &\text{ if }\ {1/2\leq x  \leq 1},
  \end{cases}
\end{align}
where $g(x):= h(x)+x-1$ and $h(x):= -x \log(x)-(1-x)\log(1-x)$ is the binary entropy.

\begin{Lmm}\label{Lmm:Bound_entropy_non-perect}
  Let $\epsilon,\varepsilon, \hat \varepsilon, \epsilon_1$ be as defined in lemma \ref{Thm:estimation}.
  If Alice is honest (but uses a non-perfect photon source), and Bob has a bounded quantum
  memory $Q$ (his quantum register $Q$ has dimension at most $D$)
  then at the end of the preparation phase, and if Alice did not abort, the smooth min-entropy of Bob on Alice string is,
  \begin{align}
    H^\epsilon_{\min}(X_1^n|QK)_\rho \geq \lambda n
  \end{align}
  with probability higher than $ 1-16(\epsilon+\varepsilon+\hat \varepsilon)-8\epsilon_1$,
   where $\lambda:= f(-D/n) -(\gamma+\alpha_4^A)-1/n$ \cite{O_F_S}, and $K$ is Bob's classical register.
\end{Lmm}
Here we have used that -- as proven in Theorem \ref{Thm:estimation} -- with
probability higher than $ 1-16(\epsilon+\varepsilon+\hat \varepsilon)-8\epsilon_1$
dishonest Bob gets at most $(\gamma  +\alpha_4^A)n$ extra bits of information
due to the leakage information on the bases used by Alice.

We can then reuse Lemma \ref{Lmm:Sec_Alice} with the only difference that we have to include the probability
that Alice emits $2$ or more photons in more than $(\gamma  + \alpha_4^A)n$ ``non-failure'' rounds.

\begin{Lmm}[Security against Dishonest Bob]\label{Lmm:Sec_Alice_np}
  Let $\epsilon,\varepsilon, \hat \varepsilon, \epsilon_1$ be as defined in Lemma \ref{Thm:estimation}.
   Let $Q$ be Bob's quantum memory such that ${\rm dim}(Q)\leq D$, and the rate
  $R$ of the code $\mathcal{C}$ be such that,
  \begin{align}
    n \geq  \frac{l+2 \log(1/2\epsilon)}{\lambda-1+R}.
  \end{align}
  If Alice is honest, then Protocol \ref{Ptol:BC_imperfect_source} (with $q=2$) either aborts or is $[3\epsilon + 16(\epsilon+\varepsilon+\hat \varepsilon)+8\epsilon_1]$-hiding.
  Note that when the honest Alice aborts she is required to output uniformly random strings, so that the security definition \ref{Def:Sec_ideal_BC}
  is \emph{also} satisfied when the protocol aborts. In fact when aborting the ideal and the real state are equal.
\end{Lmm}
\begin{proof}
  The proof is exactly the same as in Lemma \ref{Lmm:Sec_Alice}, except that we add $16(\epsilon+\varepsilon+\hat \varepsilon)+8\epsilon_1$
  to the failure probability, which corresponds to the probability that there are more than $(\gamma + \alpha_4^A)n$
  rounds where at least $2$ photons have been emitted (see Theorem \ref{Thm:estimation}), and where $\lambda$ has value given by Lemma \ref{Lmm:Bound_entropy_non-perect}.
\end{proof}

\textbf{Security for Bob:}

We will start by stating a lemma similar to Lemma \ref{Thm:exist_X}, adapted to the case
of an imperfect single photon source.

\begin{Lmm}\label{Lmm:exist_X_imperf}
When Bob is honest, at the end of the preparation phase, there exist a state $\sigma_{\bar X_1^n A \mathcal{I}}$ such that
\begin{itemize}
  \item $\sigma_{\bar X_1^n A I}=\sigma_{A\bar X_1^n I'} \otimes \tau_{I''}$
  \item $\rho_{AB} = \sigma_{A (\bar X_{\mathcal{I}}\mathcal{I})}$,
\end{itemize}
where $\tau$ denotes the maximally mixed state, $I'$ is the register encoding the set of rounds where Alice got extra information from the emission of multiple photons,
and $I''$ is the register encoding the set of rounds in $\mathcal{I}$ where Alice did not get any information. Formally the registers
$I'$ and $I''$ are such that $I'\otimes I''=\mathcal{I}$.
$\rho_{AB}$ is the real state produced by the protocol between (dishonest) Alice and Bob, and where the
registers $(A, B)$ are identified with $(A,\bar X_{\mathcal{I}}\mathcal{I})$.
\end{Lmm}

In the following we will use the same reasoning as in Lemma \ref{Lmm:sec_BC_Bob}, adapting it
to the case where the multiphoton emissions are possible.

Intuitively when Bob is honest but uses a non-perfect single photon source dishonest Alice basically knows,
for a fraction $\gamma$ of the rounds, whether they belong to $\mathcal{I}$ or not.
Using similar notations as in Lemma \ref{Lmm:sec_BC_Bob}, this knowledge will help dishonest Alice when
she will have to flip $d/2$ bits from $\bar X_1^n$. Indeed she can flip the $\approx (\gamma/2) n$
bits that she knows not to be in $\mathcal{I}$. For the $\approx d/2-(\gamma/2) n$ remaining
bits, she will flip bits that are not the $\approx (\gamma/2) n$ she knows to be
in $\mathcal{I}$.

\begin{Lmm}[see \cite{NJCKW12}]\label{Lmm:Sec_Bob_np}
    Let $\epsilon,\varepsilon,\hat \varepsilon,\epsilon_1$ be as defined in Lemma \ref{Thm:estimation}.
    Let $\alpha_1, \alpha_2$ be the same as in Lemma \ref{Lmm:sec_BC_Bob}, and $\alpha_4^B$
    as defined in Protocol \ref{Ptol:BC_imperfect_source}. Let
    $\alpha_1'':= \sqrt{\frac{\ln \epsilon^{-1}}{2(1-\gamma-\alpha_4^B)n}}$, \break
     $\alpha_1':=\min\left[1/2\,;\, \frac{\alpha_1+(1-\gamma-\alpha_4^B)\alpha_1''}{\gamma+\alpha_4^B}\right]$,
    $\alpha_3:=\sqrt{\frac{\ln \epsilon^{-1}}{2 n \big[1/2-\alpha_1'' - (1/2+\alpha_1')(\gamma+\alpha_4^B)\big]}}$.
    Let $\mathcal{C}$ be an $[n,k,d]$-code with minimum distance $d$ that satisfies,
    \begin{align}
      d &\geq 2\left[(1/2+\alpha_1')(\gamma+\alpha_4^B)+\alpha_3+\frac{(e_{\rm err}+\alpha_2)(1/2+\alpha_1)}{(1/2-\alpha_1')(1-\gamma-\alpha_4^B)}\right]n
       \underset{n \rightarrow \infty}{\sim} \left(\gamma + \frac{2e_{\rm err}}{1-\gamma}\right) n.
    \end{align}
    Then when Bob is honest, Protocol \ref{Ptol:BC_perfect_source} either aborts or is $[\epsilon+16 (\epsilon+\varepsilon+\hat \varepsilon) +8\epsilon_1]$-binding according to definition \ref{Def:Sec_ideal_BC}.
    Since when honest Bob aborts he is required to reject the opening and output a random string $\tilde C_1^l$,
    the security definition
    is automatically satisfied when honest Bob aborts the protocol.
\end{Lmm}
\begin{proof}[Proof (Sketch).]
  From Lemma \ref{Thm:estimation} we know that except with probability $16(\epsilon+\varepsilon+\hat \varepsilon)+8\epsilon_1$,
  dishonest Alice gets information on at most $(\gamma + \alpha_4^B)n$ bits.

  Except with probability $\epsilon$, at most a fraction $(1/2+\alpha_1')$ of them are not in $\mathcal{I}$, so Alice can flip them
  without Bob being able to detect this. We can compute this fraction by noticing
  that on rounds where 1 photon has been emitted (there are at least $(1-\gamma-\alpha_4^B) n$ of them), the probability
  of each of these rounds to be in $\mathcal{I}$ is $1/2$ and is independent of Alice's information.
  Therefore, by Hoeffding inequality (see Theorem \ref{Thm:Hoeffding}), the number of these rounds being in $\mathcal{I}$
  should be $\leq 1/2 + \alpha_1''$, except with probability $\epsilon$. Moreover, if the protocol does not
  abort then the \emph{total} number of rounds in $\mathcal{I}$ is $m \leq (1/2+\alpha_1)n$. Combining this
  with the fact that $1/2+\alpha_1'\leq 1$ gives the expression for $\alpha_1'$.

  At least $d/2 - (1/2+\alpha_1')(\gamma+\alpha_4^B)n$ bits remains for Alice to flip.
  However she knows that she should flip these remaining bits on the position on which she did not get any information during the
  preparation phase. There are $\geq (1-\gamma-\alpha_4^B) n$ such positions. Therefore Alice's choice of
  bit flip is equivalent to uniformly sampling without replacement $ d/2 - (1/2+\alpha_1')(\gamma+\alpha_4^B)n$ positions out
  of $\geq (1-\gamma-\alpha_4^B) n$ to estimate the number $W$ of bits that Alice chooses
  to flip while being in a position in the set $\mathcal{I}$. As for Lemma \ref{Lmm:sec_BC_Bob} this is equivalent to
  first fixing Alice's bit flip and then choosing the position that are in $\mathcal{I}$ among the $(1-\gamma-\alpha_4^B) n$
  available positions. Using Hoeffding inequality we get that,
  \begin{align}
    \Pr\Big(W < n \big[1/2-\alpha_1'' - (1/2+\alpha_1')(\gamma+\alpha_4^B)\big](d/2n - (1/2+\alpha_1')(\gamma+\alpha_4^B)-\alpha_3) \Big)\\
    \leq \exp\big(-2 n \big[1/2-\alpha_1'' - (1/2+\alpha_1')(\gamma+\alpha_4^B)\big] \alpha_3^2\big)=\epsilon.
  \end{align}

Now if
\begin{align}\label{eq:cond_d_imperf}
  d\geq 2\left[(1/2+\alpha_1')(\gamma+\alpha_4^B)+\alpha_3+\frac{(e_{\rm err}+\alpha_2)(1/2+\alpha_1)}{(1/2-\alpha_1')(1-\gamma-\alpha_4^B)}\right]n,
\end{align}
then with probability $\geq 1-\epsilon-16(\epsilon+\varepsilon+\hat \varepsilon)-8\epsilon_1$,
\begin{align}
  W &\geq n \big[1/2-\alpha_1'' - (1/2+\alpha_1')(\gamma+\alpha_4^B)\big](d/2n - (1/2+\alpha_1')(\gamma+\alpha_4^B)-\alpha_3) \\
  &\geq (e_{\rm err}+\alpha_2)(1/2+\alpha_1)n \geq (e_{\rm err}+\alpha_2)m.
\end{align}
This means that if eq.~\eqref{eq:cond_d_imperf} is satisfied there is a probability at
 most $\epsilon+16(\epsilon+\varepsilon+\hat \varepsilon)+8\epsilon_1$ that
Alice can cheat \emph{and} make Bob accept.
\end{proof}

\subsection{OT with an imperfect single photon source}

In this section we state more formally our impossibility result for a secure Oblivious
Transfer protocol. In particular we show that if a protocol satisfy Assumption \ref{Hyp:assumptions_imposs},
then Protocol \ref{Ptol:dishonest_Alice_OT} allows dishonest Alice to cheat.\\

\subsubsection{Informal description of the settings}\label{Sec:OT_intuition}

We recall that dishonest Alice's goal is to guess correctly the bit $C$ that is given
to honest Bob by the protocol (the protocol gives him an random bit $C$ and a bit string $S_C$).
In section \ref{Sec:OT_imperfect} we already give a simple example on how an attack could work
on a protocol like Protocol \ref{Ptol:OT}. Here we explain informally what is the general form of the protocols
to which our impossibility result applies.
In the next section we will make this setup definition more precise. Our impossibility result applies to
protocols of the following form.
\begin{description}
  \item[First phase] In a first phase, called the quantum phase,
  Alice and Bob can used classical and quantum communication. This phase outputs
  string $X_1^n$ to Alice and a string $X_{\mathcal{I}}$ and set of indices $\mathcal{I}$ to Bob, where
  $X_{\mathcal{I}}$ is a string formed by the bits of the string $X_1^n$ that are placed at indices in $\mathcal{I}$.
  In order to model the leakage of information due to the multiphoton emissions (see section \ref{Sec:OT_imperfect})
  we assume that Alice receives two extra sets of indices $I_G$ and $I_B$. This two sets are correlated to
  the set $\mathcal{I}$. In particular we will consider elements of $I_G$ are more likely to belong to
  $\mathcal{I}$ than elements in $I_B$.
  In the simple example of Section \ref{Sec:OT_imperfect}, dishonest Alice could compute these sets
  from the leakage information concerning the bases Bob has used in this phase. Moreover, in this specific example we
  had that $I_G\subseteq \mathcal{I}$ and $I_G \subseteq \mathcal{I}^c$, where $\mathcal{I}^c$ is the complement
  of $\mathcal{I}$.

  \item[Second phase] The second phase of the protocol is purely classical, that is they only send
  classical messages. Alice and Bob should use the data they got from the first phase in order to compute the
  desired strings $(S_0,S_1)$ and bit $C$ that the OT protocol should produce (see Definition \ref{Def:OT_informal}).
\end{description}

Note that we don't specify the specific form for the first phase, we simply require that it outputs the
strings $X_1^n$, $X_{\mathcal{I}}$ and the set $\mathcal{I}$ with some probability distribution,
as well as the extra sets $I_G$ and $I_B$ when Alice is dishonest. The strategy we use to
break security of MDI OT protocols is a semi-honest strategy. This means that Alice will
essentially run the protocol honestly\footnote{She still has full control over the measurement station. In particular
everything that Bob sends to the measurement station is considered to be in dishonest
Alice's possession.} but record all the information from the communication
between her and Bob. In particular, in the quantum phase Alice extracts -- from the quantum signal Bob
sends to the measurement station -- information about set $I_G$ and $I_B$ before
applying the measurement that the station should normally apply. Of course our attack
rely on the fact that the set $I_G$ and $I_B$ are sufficiently large so that Alice gets enough statistics
to have a good guess of Bob's bit $C$. In other words we need that Bob's photon source leaks enough information. This is captured in
the second equation of eq.~\eqref{eq:assumptions_imposs} in Assumptions \ref{Hyp:assumptions_imposs}.

After this quantum phase, we assume that Alice and Bob can post process the data they received from the quantum phase,
by using purely classical communication. Since we assume that dishonest Alice is
semi-honest, we will assume that she runs the post-processing honestly but records all the information
she receives from, or sends to Bob.

 In the post-processing of Protocol \ref{Ptol:OT} Bob chooses uniformly at
random the bit $C$, and then renames the sets $\mathcal{I}$ and $\mathcal{I}^c$ into
$I_0$ and $I_1$ in such a way that $\mathcal{I}=I_C$. Bob then sends $(I_0,I_1)$ to Alice.
The information $(I_0,I_1)$ sent by Bob to Alice, should not by itself reveal bit $C$. But because
Alice holds the extra sets $I_G$ and $I_B$, she can determine which set from $(I_0,I_1)$ corresponds
to set $\mathcal{I}$, and therefore she learns the value of bit $C$. In the general settings, we will only assume
that from all the information Alice has she can compute two sets $I_0$ and $I_1$ such that $I_C \subseteq{I}$
and $I_{1-C} \not\subseteq \mathcal{I}$. This is the second assumption in Assumptions \ref{Hyp:assumptions_imposs}.

In the following sections, we describe in details how we generalize this idea of attack to a more general settings.

\subsubsection{Settings Definition}

In this section we defined the settings in which our theorem holds. Theorem \ref{Thm:Alice_cheat},
 states that any protocol that has the form we describe below, and that is correct, and secure against Bob,
can be attacked by dishonest Alice. That is, it is always possible for Alice to correctly guess
Bob' bit $C$ with sufficiently high probability. Dishonest Alice's cheating strategy in
is given in Protocol \ref{Ptol:BC_imperfect_source}. In order to generalize the discussion of Section \ref{Sec:OT_intuition}, all the random variables mentioned in Section \ref{Sec:OT_intuition}, $X_1^n,\mathcal{I},I_0,I_1,C,\ldots$
will be redefined in a more abstract manner.

In order to prove our result we will forget about the quantum part of the OT protocol, and start directly
in a scenario, in which Alice and Bob share from the start the type of correlation they would have had
by running a preparation phase similar to Protocol \ref{Ptol:OT}.

In particular we will assume that the Preparation phase gives the following to Alice and Bob:

\begin{description}
  \item[Honest Alice] Alice gets a random bit string $X_1^n$ with probability distribution $P_{X_1^n}$.
  \item[Honest Bob] Bob gets a random subset $\mathcal{I}\subseteq [n]$ with probability distribution $P_{\mathcal{I}}$,
   and the string
  $X_\mathcal{I}$, whose bits are the bits of $X_1^n$ that are indexed by $i \in \mathcal{I}$.
\end{description}
When one of the parties is dishonest we will assume they have the following additional information as input:
\begin{description}
  \item[\textcolor{black}{Dishonest Alice}] Dishonest Alice gets the same $X_1^n$ as when she was honest, \textcolor{black}{plus}
  the sets \textcolor{black}{$I_G, I_B\subseteq [n]$}, which are sets of indices satisfying the following:\\
  $I_G \cap I_B = \emptyset$ and $|I_G \cup I_B| =\gamma n$ for some $\gamma \in ]0,1[$.\\
  $\forall i\in I_G\cup I_B$
  \begin{itemize}
    \item[-] \textbf{If} $i\in \mathcal{I}$ \textbf{then} $i \in I_G$ with probability $1/2 (1+\mu)$ or $i \in I_B$ with probability $1/2 (1-\mu)$.
    \item[-] \textbf{If} $i \notin \mathcal{I}$ \textbf{then} $i \in I_G$ with probability $1/2 (1-\mu)$ or $i \in I_B$ with probability $1/2 (1+\mu)$.
  \end{itemize}
  $\gamma$ represents the faction of rounds in which more than two photons have been emitted (we should have $\gamma \approx p_{\geq 2}/(1-p_{0})$,
  the probability that more than two photons are emitted when
  at least one is emitted). $\mu\in]0,1]$ models
  Alice's probability of guessing Bob's basis conditioned on receiving several photons from Bob.

  Note that this definition can be seen as first giving $I_G \cup I_B$ to Alice and then
   giving her $I_G$ and $I_B$ through the probabilistic process described above.

  \item[{Dishonest Bob}] When Bob is dishonest we will assume that he holds a classical register {$K$}
  and a quantum register {$Q$}
  such that his min-entropy rate $\frac{H_{\min}(X_1^n|{KQ})}{n}$ is smaller than the one of honest Bob.
\end{description}

Let $M_{BA}$ be the random variable that describes the set of the messages sent
from Bob to Alice, and $M_{AB}$ be the random variable that describes the messages sent from Alice to Bob.
 The random variable composed of these two
variables will be called $M$, in other words $M:=(M_{AB},M_{BA})$.

The output of honest Alice is $(S_0,S_1):=(f_0(X_1^n,M), f_1(X_1^n, M)) \in \{0,1\}^l\times\{0,1\}^l$, where $f_0$ and $f_1$ are two
functions. The output of honest Bob is $(C, S_C):= (g(X_{\mathcal{I}}, \mathcal{I},M), \tilde g(X_{\mathcal{I}}, \mathcal{I},M))$,
where $g$ and $\tilde g$ are two other functions. These four functions model the operations
that honest Alice and Bob have to perform according to the protocol they are running.

Before estimating Alice's cheating probability (see Theorem \ref{Thm:Alice_cheat}), we will need the following definition.

\begin{Def}
  Let $J\subseteq [n]$ be a set of indices. Let $f_0,f_1$ be the functions defined
  above. We will say that $J$ stabilizes a function $f_a$ ($a\in\{0,1\}$) with respect to (w.r.t.) random string $X_1^n$ and
  random variable $M$ when the value $(x_1^n,m)$ of random variable $(X_1^n,M)$ is such that
  $J$ stabilizes $f_a$ w.r.t.~$x_1^n$ and m.
  We will say that $J$ stabilizes the
  function $f_a$ w.r.t.~$x_1^n$ and $m$ if
   $\ x_{J^c} \mapsto f_a((x_{J^c},x_J),m)$ is constant for all $x_{J^c} \text{ s.t.}~\Pr\big((X_1^n,M)=((x_{J^c},x_J),m)\big)\neq 0$,
   where $(x_{J^c},x_J)$ denotes
   the string composed of the bits $x_J$ of  and $x_{J^c}$ at the positions corresponding to the
   sets $J$ and $J^c$.
\end{Def}

Intuitively this definition captures the notion of a function $f$ depending only on
the values of the bits of $X_1^n$ at positions indexed by the set $J\subseteq[n]$.\newline

\subsubsection{Assumptions and main Theorem}

In this section we state the assumptions we make to prove our theorem and prove Theorem \ref{Thm:Alice_cheat}.
Since we assume Alice is semi-honest her cheating strategy consists in making her guess on Bob's
bit $C$ using all the information she has collected during the protocol. Therefore we can consider that
her cheating strategy is an algorithm she runs at the end of the protocol on all her data. The cheating strategy we
use is described in Protocol \ref{Ptol:dishonest_Alice_OT}. The basic idea of the protocol is the following.
At the end of the protocol Alice has the two sets $I_0$ and $I_1$ that are correlated to bit $C$ and set $\mathcal{I}$
in the following way. If $C=0$ then $I_0\subseteq \mathcal{I}$ and $I_1\not\subseteq \mathcal{I}$. If $C=1$ the situation is reversed (see
previous section)\footnote{They have to be correlated to $C$ in this way for the OT protocol to be correct,
and secure against dishonest Bob.}. In themselves, these sets do not reveal the value of bit $C$ since Alice should not know anything
about set $\mathcal{I}$. However, since there has been information leakage during the protocol, she does know something
about set $\mathcal{I}$. She knows that indices in $I_G$ are more likely to belong to $\mathcal{I}$ than the ones
in $I_B$, and this allows her to guess with some probability which set $I_0$ or $I_1$ is a subset
of $\mathcal{I}$, and therefore it allows her to guess the value of bit $C$. Let us state more precisely the assumptions
we use to prove Theorem \ref{Thm:Alice_cheat}.

Let $\mathfrak{P}_F$ be the following statement:
``$\exists F(\cdot,\cdot)$ such that $F(X_1^n,M)=:(I_0,I_1)$ where $I_0,I_1 \subseteq [n]$ are such that
$I_{C}$ stabilizes $f_C$ but not $f_{1-C}$ (w.r.t.~$(X_1^n,M)$), and $I_{1-C}$ stabilizes $f_{1-C}$ but not
$f_C$ (w.r.t.~$(X_1^n,M)$)'', where $C:=g(X_1^n,\mathcal{I},M)$.''

 If $\mathfrak{P}_F$ is true then one can define $\alpha \in ]0,1]$ such that $|I_{1-C}\backslash I_C\cap (I_G \cup I_B) \cap \mathcal{I}| = (1-\alpha) |I_{1-C}\backslash I_C \cap (I_G \cup I_B)|$,
 \textit{i.e.}~$\alpha$ is the fraction of rounds in $I_{1-C}\backslash I_C\cap (I_G \cup I_B)$ that are not in $\mathcal{I}$.

\begin{Hyp}\label{Hyp:assumptions_imposs}
  Let $I_0,I_1,C,M,X_1^n,\mathcal{I}, I_G,I_B, \alpha$, and $\mu$  be as defined above. Let $\delta\in]0,1/2]$.
    Let $\kappa:= \min\big(\big|I_0\backslash I_{1} \cap (I_G \cup I_B)\big|; \big|I_{1}\backslash I_{0} \cap (I_G \cup I_B)\big|\big)$.
    Let $\Omega_{\kappa}$ be the event: ``$\kappa\geq 1$''. We assume in Theorem \ref{Thm:Alice_cheat} that:
    \begin{align}\label{eq:assumptions_imposs}
      \begin{cases}
       \mathfrak{P}_F \text{ is true},\\ \vspace{-5pt} \\
        \Pr(\Omega_{\kappa} ) \text{ is non-negligible in $n$},
      \end{cases}
    \end{align}
\end{Hyp}

Now we can state and prove our theorem that shows that Protocol \ref{Ptol:dishonest_Alice_OT}
is a strategy that allows dishonest Alice to to cheat.

 \begin{Thm} \label{Thm:Alice_cheat}
Let $I_0,I_1,C,M,X_1^n,\mathcal{I}, I_G,I_B, \alpha$, and $\mu$  be as defined above. Let $\delta\in]0,1/2]$.
   Let $\kappa:= \min\big(\big|I_0\backslash I_{1} \cap (I_G \cup I_B)\big|; \big|I_{1}\backslash I_{0} \cap (I_G \cup I_B)\big|\big)$.
   Let $\Omega_{\kappa}$ be the event: ``$\kappa\geq 1$''. Let $P_{\rm guess}$ be the
   maximum probability that Alice correctly guesses Bob's bit $C$.

   If Assumptions \ref{Hyp:assumptions_imposs} are satisfied by the protocol run between Alice and Bob, and if
   this protocol is correct, and secure against dishonest Bob,
   then dishonest Alice's strategy presented in Protocol \ref{Ptol:dishonest_Alice_OT} allows
   Alice to guess $C$ with probability $P_{\rm guess} = 1/2+{\rm adv}$, where ${\rm adv}$ satisfies
   \begin{align} \label{eq:cheating_advantage}
     {\rm adv}\geq \Pr\Big( \Omega_{\kappa} \Big) \times \alpha \mu.
   \end{align}
   We can also prove that $\alpha\geq 1/n$, which is not negligible in $n$.
  \end{Thm}
 \begin{proof}
   In order to prove the theorem we will lower bound Alice's guessing probability $P_{\rm guess}$, for a protocol
   satisfying Assumptions \ref{Hyp:assumptions_imposs}. In particular we want to show that $P_{\rm guess}$ is
   larger than $1/2$ by a non-negligible amount. Before doing that let us spell out
   important consequences of a protocol being correct and secure against Bob.

   Because we assume that $\mathfrak{P}_{F}$ is true, the sets
   $(I_0,I_1):=F(X_1^n,M)$ are well defined.
   In order to get correctness
   we should have that $I_C\subseteq I$, and for having security against Bob
   it is necessary that $I_{1-C}\nsubseteq I$, where $C$ is the bit held by
   honest Bob that Alice tries to guess. Let us call $b$ the bit that corresponds to
   dishonest Alice's guess of Bob's bit $C$. We can then write,

   \begin{align}
     P_{\rm guess} = \Pr(b=C) &= \Pr(\Omega_\kappa) \Pr(b=C | \Omega_\kappa) + (1-\Pr(\Omega_\kappa)) \Pr(b=C|\neg \Omega_\kappa),\\
     &\geq \Pr(\Omega_\kappa) \Pr(b=C | \Omega_\kappa) + (1-\Pr(\Omega_\kappa)) 1/2.
   \end{align}

   From Assumptions \ref{Hyp:assumptions_imposs} we have that $ \Pr(\Omega_\kappa)$ is not negligible.
   Intuitively, saying that $ \Pr(\Omega_\kappa)$ is not negligible ensures that there has been
   information leakage during the quantum phase of the protocol. If $\Pr(\Omega_\kappa)$ were negligible we already
   know by Theorem \ref{Thm:Sec_OT_perf} that a protocol like Protocol \ref{Ptol:OT} would be secure.
   As a consequence, we will focus on computing $\Pr(b=C | \Omega_\kappa)$.

   In Protocol \ref{Ptol:dishonest_Alice_OT},
   Alice chose uniformly at random an index $i_r\in I_r|I_{1-r} \cap (I_G \cup I_B)$ and check whether
   $i_r$ ends up in $I_G$ or $I_B$. The idea is that if $r=C$ the probability that $i_r$ ends up
    in $I_G$ is slightly higher than the one of ending up in $I_B$. If $r=1-C$ it biased towards ending up
    in $I_B$. Therefore, if she outputs $b=r$ when $i_r \in I_G$, and outputs $b=1-r$ if $i_r \in I_B$ she will
    have a probability of guessing correctly bit $C$ slightly higher than $1/2$,
    which is what we are trying to prove.

   Note that the event $\Omega_{\kappa}$ depends on the ``value'' of the set
   $I_G\cup I_B$, but is completely independent on how $I_G\cup I_B$ is partitioned
   into the sets $I_G$ and $I_B$.
   In particular the probability for a round in $I_G\cup I_B$ to be in $I_G$ is independent
   of $\Omega_\kappa$.

 Let us now write Alice's guessing probability conditioned on $\Omega_\kappa$, with $r$ and $i_r$ as defined by Protocol \ref{Ptol:dishonest_Alice_OT}:
 \begin{align}
   P_{{\rm guess}|\Omega_\kappa} &= \Pr(b=C|\Omega_\kappa) \\
   &= \Pr(r=C|\Omega_\kappa) \Pr(i_r \in I_G|r=C,\Omega_\kappa) + \Pr(r=1-C|\Omega_\kappa) \Pr(i_r \in I_B | r=1-C,\Omega_\kappa),
 \end{align}
 where $r$ is a uniformly random bit chosen by Alice in Protocol \ref{Ptol:dishonest_Alice_OT}.
 As a consequence, $\Pr(r=C|\Omega_\kappa)= \Pr(r=1-C|\Omega_\kappa)=1/2$.
 From the definition of $I_G$ and $I_B$, and their independence from $\Omega_\kappa$ we get that,
 \begin{align}
   \Pr(i_r \in I_G|r=C,\Omega_\kappa)&= \Pr(i_r \in I_G|r=C)=1/2(1+\mu),\\
   \Pr(i_r \in I_B | r=1-C,\Omega_\kappa) &= \Pr(i_r \in I_B | r=1-C)=1- 1/2\alpha(1-\mu) -1/2(1-\alpha)(1+\mu).
 \end{align}
 Plugging this into the expression for $P_{{\rm guess}|\Omega_\kappa}$ we get that:
 \begin{align}
   P_{{\rm guess}|\Omega_\kappa}=1/2\big(1/2(1+\mu) + 1- 1/2\alpha(1-\mu) -1/2(1-\alpha)(1+\mu)\big)=1/2(1+\alpha \mu).
 \end{align}
As expected the probability that Alice correctly guesses the value of bit $C$ is a bit higher than $1/2$.

 Combining this with the fact that $\kappa\geq 1$ is true with probability $\Pr(\Omega_\kappa)$ leads to eq.~\eqref{eq:cheating_advantage}.
 That is, Alice's overall probability of guessing correctly bit $C$ is still slightly
 higher than $1/2$, namely it is higher than, \[1/2 + \Pr(\Omega_\kappa) \times \alpha \mu.\]

 As we stated earlier $I_{1-C} \nsubseteq \mathcal{I}$, meaning that at least one index in $I_{1-C}$ is not in $\mathcal{I}$, and
 since $I_{1-C}\backslash I_C \cap (I_G\cup I_B)$ cannot be larger than the total length of the string $X_1^n$ (which is obviously $n$), we must have $\alpha \geq 1/n$.

 \end{proof}

 \section{Acknowledgment}
 JR and SW are supported by NWO VIDI, and ERC
 Starting Grant and NWO Zwaartekracht QSC. We would like to thank Victoria Lipinska, Mark Steudtner, Kenneth Goodenough, Kaushik Chakraborty and
 Bas Dirkse for giving useful comments on this manuscript.

 \bibliography{bib_article}
 \bibliographystyle{alpha}

\appendix

\section{Why doesn't dishonest Bob get any advantage by selectively discarding rounds when Alice uses a perfect single photon source?}
\label{Sec:App_selc_discard}

In this section we explain why for our proof we can consider that we can simply evaluate
the min-entropy bound of Lemma \ref{Lmm:min-entropy_bound} as if Bob were honest in
choosing which rounds he announces to be lost. In other words we explain why dishonest
Bob can't get any advantage by selectively discarding rounds.

In Protocol \ref{Ptol:BC_perfect_source}, Alice sends $n'$ BB84 states to Bob using a perfect single photon source.
This, by purification of the states she sends, is equivalent as to Alice preparing
$n'$ EPR pairs, and sending half of each pairs to dishonest Bob, and randomly measuring her
halfs of EPR pairs in the $X$ or $Z$ basis. This allows us to
delay Alice's measurements to the end of the preparation phase.

The bound we use for the min-entropy is independent of the details of the state.
Indeed the bound works as follows. For \emph{any} state $\rho_{A^{n} E}$ (for some $n\in \mathbb{N}$), if Alice's measurements
(modeled by the CPTP map $\mathcal{M}_{A_1^n \mapsto X_1^n}$) on the
systems $A_1^n$ (outputting bit string $X_1^n$) satisfy some condition (that is indeed satisfied
when Alice randomly measures in the $X$ or $Z$ basis \cite{O_F_S}), then
$H_{\min}^{\epsilon}(X_1^n | E)_{\mathcal{M}(\rho)} \geq B(H_{\min}(A_1^n | E)_{\rho}/n) \cdot n$,
where $B(\cdot)$ is some function that bounds the min-entropy rate.

Since the bounds applies to \emph{any} state, one can then choose $\rho_{A_1^n E}$ to
 be the state of the protocol after that Bob (holding register $E=KQ$, where $K$ is
  classical and $Q$ is the quantum state in his memory) has stored quantum information and after
he has announced which rounds are kept and which are not, but before Alice has measured.
Using the bounded storage assumption ($\log {\rm dim}(Q)\leq D$) we can bound $H_{\min}(A_1^n | E)_{\rho}\geq -D$.
This leads us to $H_{\min}^{\epsilon}(X_1^n | E)_{\mathcal{M}(\rho)} \geq B(-D/n)\cdot n$ as stated in Lemma \ref{Lmm:min-entropy_bound}.
Note that this bound is evaluated on the state conditioned on Bob keeping some particular rounds,
but the bound does not depend on the strategy he uses for choosing which rounds he keeps and which he discards.

For Protocol \ref{Ptol:OT} the same reasoning apply. Indeed even though we use a different bound,
the bound we use is also independent of the details states on which the entropy is evaluated.

\section{Proof of Lemma \ref{Thm:estimation}}
\label{Sec:Thm:estimation}

In this section we will explain how the honest party $H \in \{A,B\}$ can use the decoy states in order to estimate
a lower-bound $L_{H1}$ on $n_1^H$. To do so we will use techniques inspired by \cite{CXC14}.
In the following we will detail the analysis considering that Alice is honest. The case when Bob is honest follows the same structure.

First we can observe that Protocol \ref{Ptol:BC_imperfect_source}
is equivalent to a virtual protocol where Alice first chooses the number $k$ of photons
she is sending according to a probability distribution $p_k$, and the encoding basis with
probability $p_{\theta}$, and only after the station
reveals the measurement outcome $o$ she chooses the signal intensity $a\in \{a_s,a_{d_1}\ldots a_{d_q}\}$
according to probability distribution $p_{a|k}$ (this choice in independent from $\theta$ and outcome $o$).
The probability distribution $p_{k}$ and $p_{a|k}$
in the virtual protocol can be deduced from the distribution $p_a$, and $p_{k|a}$ of Protocol \ref{Ptol:BC_imperfect_source}
via Bayes' rule.

As a consequence for any set $S_{k,o,\theta}^A$  of rounds where Alice has emitted $k$ photons encoded in the
basis $\theta$ ($\theta=0$ for the standard basis, and $\theta=1$ for the Hadamard basis)
and the measurement station (or dishonest Bob) reported measurement
outcome $o$ (with $o\neq$ failure), each subset of $S_{k,o,\theta}^A$ corresponding
to intensity $a$ can be seen as a random sample of $S_{k,o,\theta}^A$.
Therefore we can use (classical) random sampling theory to estimate $L_{A1}$,
like Chernoff's bound for example. In particular we will use the following lemma proven in Ref.~\cite{CXC14},

\begin{Lmm}\label{Lmm:modified_Chernoff}
  Let $X_1,\ldots, X_n$ be $n$ independent Bernoulli random variables such that $\Pr(X_i=1)=p_i$, and let
  $X:=\sum_i X_i$ and $\zeta:=\mathbb{E}(X)=\sum_i p_i$. Let $x$ be the observed outcome of $X$ for a certain trial and
  $\Gamma:=x-\sqrt{n/2 \ln(1/\epsilon)}$ for a certain $\epsilon>0$. If $\varepsilon, \hat \varepsilon >0$ are such
  that $(2\varepsilon^{-1})^{1/\zeta_L} \leq \exp(3/(4\sqrt{2}))^2$ and $(\hat \varepsilon^{-1})^{1/\zeta_L}<\exp(1/3)$ then
  $x$ satisfies,
  \begin{align}
    x=\zeta+\delta,
  \end{align}
  except with probability $\epsilon+\varepsilon+\hat \varepsilon$, where $\delta \in [-\Delta, \hat \Delta]$,
  with $\Delta:= g(x,\varepsilon^4/16),\, \hat \Delta := g(x, \hat \varepsilon ^{3/2})$ and $g(x,y):= \sqrt{2x \ln(y^{-1})}$.
  Here $\varepsilon (\hat \varepsilon)$ denotes the probability that $x<\zeta-\Delta\, (x>\zeta+\hat \Delta)$.
\end{Lmm}

This lemma is a variation of the Chernoff's bound, where the bounds on the fluctuations $\Delta (\hat \Delta)$ do not depend
on the expectation value $\zeta:=\bE(X)$ of the random variable $X$, but only on the observed value $x$ of $X$ (and the epsilons).\\

Let $S_{k,o,\theta}^A$ be the set of rounds as defined above, and let $X_{i|k,o,\theta}^{a}$
be $1$ if the $i^{\rm th}$ element of $S_{k,o,\theta}^A$ corresponds to an emission of a state (from honest Alice) with
intensity $a$, and $0$ otherwise. Let
\begin{align}
  X_{o,\theta}^a = \sum_{k} \sum_{i=1}^{|S_{k,o,\theta}^A|} X_{i|o,k,\theta}^A,
\end{align}
with $\zeta_{o,\theta}^a:=\mathbb{E}(X_{o,\theta}^a)=\sum_{k} p_{a|k}\ |S_{k,o,\theta}^A|$. Let $x_{o,\theta}^a$ be an observed
outcome of $X_{\theta,o}^a$. Then applying Lemma \ref{Lmm:modified_Chernoff} we have that for some
$(2\varepsilon^{-1})^{1/\Gamma_{o,\theta}^a} \leq \exp(3/(4\sqrt{2}))^2$, $(\hat \varepsilon^{-1})^{1/\Gamma_{o,\theta}^a}<\exp(1/3)$ with
\begin{align}
  \Gamma_{o,\theta}^a=x_{o,\theta}^a - \sqrt{\sum_a x_{o,\theta}^a /2 \ln(1/\epsilon)},
\end{align}
the following must be satisfied:
\begin{align}\label{eq:constrant_opt}
  x_{o,\theta}^a = \sum_{k} p_{a|k} |S_{k,o,\theta}^A| + \delta_{a,o,\theta},
\end{align}
except with probability $\epsilon+\varepsilon+\hat \varepsilon$, where $\delta_{a,o,\theta}\in [\Delta_{a,o,\theta},\hat \Delta_{a,o,\theta}]$,
with $\Delta_{a,o,\theta}=g(x_{o,\theta}^a, \varepsilon^4/16)$ and $\hat \Delta_{a,o,\theta} =g(x_{o,\theta}^a, \hat \varepsilon^{3/2})$.

Since $n_1^A= \sum_{o,\theta} n_{1|o,\theta}^A$ it is enough to find a lower bound on
$n_{1|o,\theta}^A$ for all values of $(o,\theta)$ in order to find a lower bound $L_{A1}$ on $n_1^A$. Then using concentration bounds
one can write that for each value of $(o,\theta)$
\begin{align}
  n_{1|o,\theta}^A \geq p_{a_s|k=1} |S_{1,o,\theta}^A| - g(p_{a_s|k=1} |S_{1,o,\theta}^A|, \epsilon_1),
\end{align}
except with probability $\epsilon_1$.
For a fixed value of $(o,\theta)$ one
can find a lower-bound on $|S_{1,o,\theta}^{A}|$ by minimizing
$|S_{1,o,\theta}^{A}|$ under the constraints given by eq.~\eqref{eq:constrant_opt}.
This can be solved by using linear programming \cite{Vanderbei}, or we can use a simplified version of this reasoning to
find analytical (but looser) bounds. This is what we will be doing in the following section.

\subsection{Simple Analytical Bound}

\newcommand{\myangle}{40}

\begin{figure}[h]
  \center
  \begin{tikzpicture}[scale=2]
    \draw[step=.5cm, gray, very thin] (-2.2,-2.2) grid (2.2,2.2);
    \draw[->,thick] (-2.25,0) -- (2.25,0) node[below] (x axis){$|S_{1,o,\theta}^A|$};
    \draw[->,thick] (-2,-2.25) -- (-2,2.25) node[right] (y axis){$|S_{\geq 2,o,\theta}^A|$};
    \draw[fill=red] (\myangle:-1) -- (0:1) -- (\myangle:1) -- (0:-1) -- cycle;
    \node[] (p1) at (\myangle:-1){};
    \node[] (p2) at (\myangle:1){};
    \node[] (p3) at (0:-1){};
    \node[] (p4) at (0:1) {};
    \draw[-] (p4) to +(90+\myangle/2:2);
    \draw[-] (p4) to +(90+\myangle/2:-2);
    \draw[-] (p1) to +(90+\myangle/2:2);
    \draw[-] (p1) to +(90+\myangle/2:-1.5);
    \draw[-] (p1) to  +(\myangle/2:2.5);
    \draw[-] (p1) to  +(\myangle/2:-1);
    \draw[-] (p3) to  +(\myangle/2:2.5);
    \draw[-] (p3) to  +(\myangle/2:-1);
    \node[rounded corners=3pt,fill=white,opacity =0.8] at (p1){$p_1$};
    \node[rounded corners=3pt,fill=white,opacity =0.8] at (p2){$p_2$};
    \node[rounded corners=3pt,fill=white,opacity =0.8] at (p3){$p_3$};
    \node[rounded corners=3pt,fill=white,opacity =0.8] at (p4){$p_4$};
  \end{tikzpicture}
  \captionof{figure}{Each of the line is defined by one of the four inequalities in \eqref{eq:system}. The red region
  is the set of points that satisfies the four linear constraints from \eqref{eq:system}. Since we are
  optimizing a linear function with linear constraints, by linear programming we know that the optimum
  is reached for one of the four extreme points $p_1,p_2,p_3,p_4$, which are at the intersection of the lines. In the
  particular case of this figure $\min |S_{1,o,\theta}^A|$ is reached in $p_3$.}
  \label{Fig:cons_region}
\end{figure}
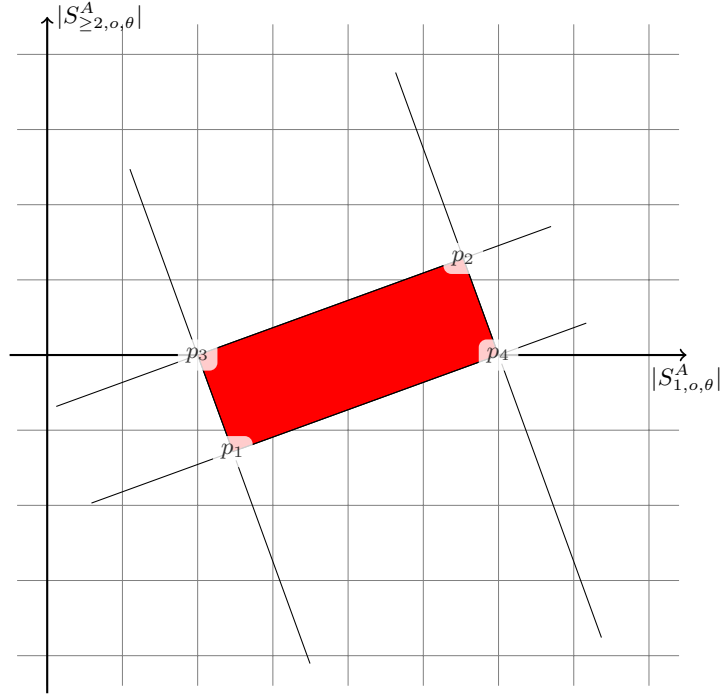

In this section we propose to find a simple analytical bound on $n_1^A$, using the reasoning and
methods of the previous section. To do so we will minimize $|S_{1,o,\theta}^A|$ for a fixed value
for $(o,\theta)$. Moreover we will restrict ourselves to the use of only $2$ decoy states and
one signal state, \textit{i.e.}~$a\in\{a_s,a_{d_1},a_{d_2}\}$.

In the previous section we have split the rounds into many sets $S_{k,o,\theta}$ ($1$ set for each value of $k$).
Here we split the round into two sets $S_{1,o,\theta}^A$ and $S_{\geq 2,o,\theta}^A$.

With this in mind we can rewrite equation \eqref{eq:constrant_opt} as the following system of inequalities,
\begin{align}\label{eq:system}
  \begin{cases}
    x_{o,\theta}^{a_{d_1}} + \Delta_{a_{d_1},o,\theta}\geq p_{a_{d_1}|k=1} \cdot |S_{1,o,\theta}^A| + p_{a_{d_1}|k\geq 2} \cdot |S_{\geq 2,o,\theta}^A|  \\
    x_{o,\theta}^{a_{d_1}} -\hat \Delta_{a_{d_1},o,\theta}\leq p_{a_{d_1}|k=1} \cdot |S_{1,o,\theta}^A| + p_{a_{d_1}|k\geq 2} \cdot |S_{\geq 2,o,\theta}^A|  \\
    x_{o,\theta}^{a_{d_2}} +\Delta_{a_{d_2},o,\theta}\geq p_{a_{d_2}|k=1} \cdot |S_{1,o,\theta}^A| + p_{a_{d_2}|k\geq 2} \cdot |S_{\geq 2,o,\theta}^A| \\
    x_{o,\theta}^{a_{d_2}} - \hat \Delta_{a_{d_2},o,\theta}\leq p_{a_{d_2}|k=1} \cdot |S_{1,o,\theta}^A| + p_{a_{d_2}|k\geq 2} \cdot |S_{\geq 2,o,\theta}^A|
  \end{cases}
\end{align}

Each of the four inequalities represents half a space delimited by a straight line in $\mathbb{R}^2$. The two first
inequalities define a region delimited by two parallel
lines, and the two last inequalities define another region delimited by two other parallel lines. The set of four
inequalities is then the intersection of these two regions, see Fig.~\ref{Fig:cons_region}. Since we
are optimizing a linear function with linear constraints the minimum is reached for one of the extreme points of this region.
Each of these points corresponds to the solution of the system of equations formed by two of the inequalities
 from \eqref{eq:system} (one for decoy state $1$
 and one for decoy state $2$) by changing symbols $\leq,\geq$ into $=$. Since there are two equations for each decoy
 state, the number of extreme points must be $4$. They can be found analytically by solving this system
 of equations. In the end the lower-bound $L_{A1}$ is given by,
 \begin{align}
   L_{A1}= \sum_{o, \theta} \left[p_{a_s|k=1}\ |S_{1,o,\theta}|_{\min} - g(p_{a_s|k=1}\ |S_{1,o,\theta}|_{\min},\epsilon_1)\right],
 \end{align}

 where $|S_{1,o,\theta}|_{\min}$ is given by,
 \begin{align}
   |S_{1,o,\theta}|_{\min} = \min(V_1,V_2,V_3,V_4),
 \end{align}
 with
 \begin{align}
   V_1=\frac{p_{a_{d_1}|k\geq 2}(x_{o,\theta}^{a_{d_2}} +\Delta_{a_{d_2},o,\theta})- p_{a_{d_2}|k\geq 2} (x_{o,\theta}^{a_{d_1}} + \Delta_{a_{d_1},o,\theta})}{p_{a_{d_1}|k=1} p_{a_{d_2}|k\geq 2} - p_{a_{d_1}|k\geq 2} p_{a_{d_2}|k=1} }\\
   V_2=\frac{p_{a_{d_1}|k\geq 2}(x_{o,\theta}^{a_{d_2}} - \hat \Delta_{a_{d_2},o,\theta})- p_{a_{d_2}|k\geq 2} (x_{o,\theta}^{a_{d_1}} + \Delta_{a_{d_1},o,\theta})}{p_{a_{d_1}|k=1} p_{a_{d_2}|k\geq 2} - p_{a_{d_1}|k\geq 2} p_{a_{d_2}|k=1} }\\
   V_3=\frac{p_{a_{d_1}|k\geq 2}(x_{o,\theta}^{a_{d_2}} +\Delta_{a_{d_2},o,\theta})- p_{a_{d_2}|k\geq 2} (x_{o,\theta}^{a_{d_1}} - \hat \Delta_{a_{d_1},o,\theta})}{p_{a_{d_1}|k=1} p_{a_{d_2}|k\geq 2} - p_{a_{d_1}|k\geq 2} p_{a_{d_2}|k=1} }\\
   V_4=\frac{p_{a_{d_1}|k\geq 2}(x_{o,\theta}^{a_{d_2}} -\hat\Delta_{a_{d_2},o,\theta})- p_{a_{d_2}|k\geq 2} (x_{o,\theta}^{a_{d_1}} -\hat\Delta_{a_{d_1},o,\theta})}{p_{a_{d_1}|k=1} p_{a_{d_2}|k\geq 2} - p_{a_{d_1}|k\geq 2} p_{a_{d_2}|k=1} } .
 \end{align}

\section{Formal Security Definitions for OT and BC}\label{Sec:Definitions}
In this section you can find the formal definitions for Randomized String Commitment
and for Randomized 1-out-2 $(l,\epsilon)$-Oblivious String Transfer. These definitions
come directly from Refs.~\cite{Steph_1}.

\begin{Rmk}[on the abort events]
  The careful reader will see that the definitions below do not mention any abort event.
  In fact our protocols specify the action a party has to take when he wants to abort.
  In particular we ask the aborting party to output uniformly random outcomes, so that
  even when aborting the security definitions are satisfied.
\end{Rmk}

\begin{Def}[Randomized String Commitment]\label{Def:Sec_ideal_BC}
  Let $\tau_{R}$ denote the maximally mixed state on a register $R$.

  An $(l,\epsilon)$-Randomized String commitment scheme is a protocol between Alice and Bob that satisfies
  the following three properties.
  \begin{description}
    \item[Correctness] When both parties are honest, then there exists a state $\sigma_{C_1^l C_1^l F}$,
    called the ideal state that is defined as:
    \begin{itemize}
      \item $\sigma_{C_1^l F}:= \tau_{C_1^l} \otimes \ketbra{accept}{accept}_F$,
      \item The real state produced by the protocol $\rho_{C_1^l \tilde C_1^l F}$ is $\epsilon$-close
      to the ideal state $\sigma_{C_1^l C_1^l F}$,
      \[\rho_{C_1^l \tilde C_1^l F} \approx_\epsilon \sigma_{C_1^l C_1^l F}. \]
    \end{itemize}
    \item[Security for Alice (against dishonest Bob)] When Alice is honest, Bob is
    ignorant about $C_1^l$ before the Open phase:
    \begin{align*}
      \rho_{C_1^l B} \approx_\epsilon \tau_{C_1^l} \otimes \rho_{B}.
    \end{align*}
    The protocol is then said to be $\epsilon$-hiding.
    \item[Security for Bob (against dishonest Alice)] After the Commit
    phase and before the Open phase, there exists an ideal state
    $\sigma_{C_1^l A B}$ such that for any Open algorithm, describe by the CPTP maps
    $\mathcal{O_{AB}}$, in which Bob is honest, we have:
    \begin{itemize}
      \item Bob almost never accepts $\tilde C_1^l \neq C_1^l$:\\
        \hspace{1cm} for $(\id_{C_1^l}\otimes \mathcal{O}_{AB})(\sigma_{C_1^l AB})$ we have
        $\Pr(\tilde C_1^l \neq C_1^l \text{and } F=accept)\leq \epsilon$.
      \item The real state produced by the commitment phase is close to the ideal state:
      \begin{align*}
        \rho_{AB} \approx_\epsilon \sigma_{AB}.
      \end{align*}
      The protocol is then said to be $\epsilon$-binding.
    \end{itemize}

  \end{description}
\end{Def}
\vspace{0.5cm}

\begin{Def}[Randomized 1-out-2 $(l,\epsilon)$-Oblivious String Transfer (OST)]\label{Def:Sec_ideal_OT}\hfill \\
  Let $\tau_{R}$ denote the maximally mixed state on register $R$.

  A fully randomized 1-out-2 $(l,\epsilon)$-Oblivious String Transfer scheme is a protocol between two parties, Alice and Bob,
  that satisfies the following three conditions.
  \begin{description}
    \item[Correctness] If both parties are honest there exists an ideal state $\sigma_{S_0S_1 C S_C}$, where $S_1,S_1 \in \{0,1\}^l$ and
    $C\in\{0,1\}$, such that:
    \begin{itemize}
      \item The distribution over $S_0,S_1$ and $C$ is uniform:
      \begin{align}
        \sigma_{S_0S_1C} = \tau_{S_0}\otimes \tau_{S_1} \otimes \tau_{C}
      \end{align}
      \item The real state $\rho$ produced by the protocol is $\epsilon$-close to the ideal state:
      \begin{align}
        \rho_{S_0S_1C \hat S_C} \approx_{\epsilon} \sigma_{S_0S_1CS_C}
      \end{align}
    \end{itemize}
    \item[Security for Bob] If Bob is honest, there exists an ideal state $\sigma_{A S_0 S_1 C}$ such that:
    \begin{itemize}
      \item Alice is ignorant about $C$:
      \begin{align}
        \sigma_{A S_0 S_1 C} = \sigma_{A S_0 S_1} \otimes \tau_{C}.
      \end{align}
      \item The real state $\rho$ produced by the protocol is close to the ideal state:
      \begin{align}
        \rho_{A C \hat S_C} \approx_\epsilon \sigma_{A C S_C}
      \end{align}
    \end{itemize}
    \item[Security for Alice] If Alice is honest, there exists an ideal state $\sigma_{S_0S_1 B C}$ such that:
    \begin{itemize}
      \item Bob is ignorant about $S_{1-C}$:
      \begin{align}
        \sigma_{S_0S_1 B C} = \sigma_{S_C B C} \otimes \tau_{S_{1-C}}.
      \end{align}
      \item The real state $\rho$ is close to the ideal state:
      \begin{align}
        \rho_{S_0 S_1 B} \approx_\epsilon   \sigma_{S_0 S_1 B}.
      \end{align}
    \end{itemize}
  \end{description}

\end{Def}

\end{document}